\documentclass[pre,twocolumn,amsmath,amssymb,nofootinbib,floatfix,superscriptaddress]{revtex4}

\usepackage{enumerate}
\usepackage{graphicx,bm}
\usepackage[normalem]{ulem} 

\makeatletter
\def\graphicscale{\twocolumn@sw{0.3}{0.4}}
\def\graphicthreescale{\twocolumn@sw{0.3}{0.4}}

\begin{document}

\title{The Coulomb-Higgs phase transition of three-dimensional lattice
  Abelian-Higgs \\ gauge models  with noncompact gauge variables and gauge
  fixing}

\author{Claudio Bonati} 
\affiliation{Dipartimento di Fisica dell'Universit\`a di Pisa
        and INFN, Largo Pontecorvo 3, I-56127 Pisa, Italy}

\author{Andrea Pelissetto}
\affiliation{Dipartimento di Fisica dell'Universit\`a di Roma Sapienza
        and INFN Sezione di Roma I, I-00185 Roma, Italy}

\author{Ettore Vicari} 
\affiliation{Dipartimento di Fisica dell'Universit\`a di Pisa,
        Largo Pontecorvo 3, I-56127 Pisa, Italy}

\date{\today}

\begin{abstract}
We study the critical behavior of three-dimensional (3D) lattice
Abelian-Higgs (AH) gauge models with noncompact gauge variables and
multicomponent complex scalar fields, along the transition line
between the Coulomb and Higgs phases.  Previous works that focused on
gauge-invariant correlations provided evidence that, for a
sufficiently large number of scalar components, these transitions are
continuous and associated with the stable charged fixed point of the
renormalization-group flow of the 3D AH field theory (scalar
electrodynamics), in which charged scalar matter is minimally coupled
with an electromagnetic field.  Here we extend these studies by
considering gauge-dependent correlations of the gauge and matter
fields, in the presence of two different gauge fixings, the Lorenz and
the axial gauge fixing. Our results for $N=25$ are definitely
consistent with the predictions of the AH field theory and therefore
provide additional evidence for the characterization of the 3D AH
transitions along the Coulomb-Higgs line as charged transitions in the
AH field-theory universality class. Moreover, our results give
additional insights on the role of the gauge fixing at charged
transitions.  In particular, we show that scalar correlations are
critical only if a hard Lorenz gauge fixing is imposed.
\end{abstract}

\maketitle


\section{Introduction}
\label{intro}

Three-dimensional (3D) Abelian gauge models with charged scalar
fields, such as the scalar electrodynamics or the Abelian-Higgs (AH)
model, provide effective field theories for many emerging phenomena in
condensed-matter systems~\cite{Anderson-book,Wen-book}. For instance,
they describe low-energy collective phenomena in superconductors,
superfluids, quantum SU($N$) antiferromagnets~\cite{RS-90, TIM-05,
  TIM-06, Kaul-12, KS-12, BMK-13, NCSOS-15, WNMXS-17,Sachdev-19},
transitions between the N\'eel and the valence-bond-solid phases in
two-dimensional antiferromagnetic SU(2) quantum
systems~\cite{Sandvik-07, MK-08, JNCW-08, Sandvik-10, HSOMLWTK-13,
  CHDKPS-13, PDA-13, SGS-16}, the last system being the paradigmatic
model for the so-called deconfined quantum
criticality~\cite{SBSVF-04}.  Classical and quantum systems that have
an effective description in terms of an Abelian gauge model with
scalar fields have been extensively studied with the purpose of
identifying their quantum and thermal phases and the nature of their
phase transitions, see, e.g.,
Refs.~\cite{HLM-74,FS-79,FMS-81,DHMNP-81,DH-81,CC-82,BF-83,FM-83,
  KK-85,KK-86,BN-86,BN-86-b,BN-87,RS-90,MS-90,KKS-94,HT-96,FH-96,YKK-96,
  BFLLW-96,OT-98,KKLP-98,CN-99,KNS-02,MHS-02,
  SSSNH-02,SSNHS-03,MZ-03,NRR-03,MV-04,SBSVF-04, NSSS-04, SSS-04,
  WBJSS-05, CFIS-05, TIM-05, TIM-06, CIS-06, KPST-06, Sandvik-07,
  WBJS-08, MK-08, JNCW-08, MV-08, KMPST-08-a, KMPST-08, CAP-08, KS-08,
  ODHIM-09, LSK-09, CGTAB-09, CA-10, BDA-10, Sandvik-10, Kaul-12,
  KS-12, BMK-13, HBBS-13, Bartosch-13, HSOMLWTK-13, CHDKPS-13, PDA-13,
  BS-13, NCSOS-15, NSCOS-15, SP-15, SGS-16,WNMXS-17, FH-17, PV-19-CP,
  IZMHS-19, PV-19-AH3d, SN-19, Sachdev-19, PV-20-largeNCP, SZ-20,
  PV-20-mfcp, BPV-20-hcAH,BPV-21-ncAH,BPV-22-mpf, BPV-22, BPV-23,
  Herbut-book}.

Lattice AH models with Abelian gauge invariance and SU($N$)
global symmetry are the lattice counterparts of the multicomponent 
AH field theory, defined by the Lagrangian
\begin{equation}
{\cal L} = 
\frac{1}{4 g^2} \,F_{\mu\nu}^2+
|D_\mu{\bm\Phi}|^2
+ r\, {\bm \Phi}^*\cdot{\bm \Phi} + 
u \,({\bm \Phi}^*\cdot{\bm \Phi})^2,
\label{AHFT}
\end{equation}
where ${\bm \Phi}({\bm x})$ is an $N$-component complex scalar field,
which is minimally coupled with the electromagnetic field $A_\mu({\bm
  x})$, with $F_{\mu\nu}\equiv \partial_\mu A_\nu - \partial_\nu
A_\mu$, and $D_\mu \equiv \partial_\mu + i A_\mu$.
Renormalization-group (RG) computations, based on perturbative
expansions~\cite{HLM-74,FH-96,IZMHS-19}, functional RG
equations~\cite{FH-17}, and on the large-$N$
expansion~\cite{HLM-74,DHMNP-81,YKK-96,MZ-03,KS-08}, have shown that
the RG flow of the AH field theory has a stable charged fixed point
(CFP) for a sufficiently large number $N$ of components, i.e., for
$N>N^{\star}$, where $N^{\star}$ depends on the space dimension:
$N^\star = 7(2)$ in three dimensions~\cite{BPV-21-ncAH}. This stable
CFP is expected to represent the universality class of continuous
transitions in strongly-interacting systems with local Abelian and
global SU($N$) symmetry~\cite{BPV-22,BPV-23}, where both scalar and
gauge correlations become critical.  Critical behaviors compatible
with the universality classes of the AH field theory have been
observed in the lattice AH model with noncompact gauge fields (here
the gauge group is the additive group ${\mathbb R}$ of real numbers)
along the transition line separating the Coulomb and Higgs
phases~\cite{BPV-21-ncAH}, and in the lattice AH model with compact
gauge fields (here U(1) is the compact gauge group) and scalar fields
of charge $q\ge 2$, at the transitions between the confined and
deconfined phases~\cite{BPV-20-hcAH, BPV-22}. They have also been
observed in a noncompact model with discrete Abelian gauge group (the
Abelian group ${\mathbb R}$ is here replaced by ${\mathbb Z}$)
\cite{BF-22}, but not in the corresponding compact model \cite{BT-22}.

Numerical studies of the lattice AH model have mostly focused on the
analysis of the critical behavior of gauge-invariant observables, such
as the correlations of local bilinears of the scalar field ${\bm
  \Phi}$. The numerical estimates of the corresponding critical
exponents are in good agreement with the 3D large-$N$ field theory
results, providing a robust evidence of the existence of a statistical
critical regime along the Coulomb-Higgs (CH) transition line where the
AH field theory (\ref{AHFT}) is realized~\cite{BPV-21-ncAH,
  BPV-20-hcAH, BPV-22}.

In this work we extend the study, considering gauge-dependent
correlations of the gauge and scalar fields.  For this purpose, we add
a gauge-fixing constraint, that makes gauge-dependent correlations of
local fields well defined.  In lattice models with noncompact gauge
fields, the gauge fixing plays a crucial role. Indeed, only in the
presence of a gauge fixing, the partition function of a finite-volume
system and nongauge-invariant gauge-field correlations are
finite. This is at variance with what happens in compact formulations,
in which a gauge fixing is not necessary, and correlations of
gauge-dependent quantities are either trivial or equivalent to
gauge-invariant observables, obtained by averaging the gauge-dependent
quantity over the (compact) group of the gauge transformations
\cite{Elitzur:1975im, DeAngelis:1977su, IZ_book}. The latter
equivalence does not hold in noncompact formulations, since the group
of gauge transformations is not compact and therefore, averages over
all gauge transformations are not defined.

The use of gauge fixings in a nonperturbative context offers the
possibility of investigating whether different gauge conditions give
rise to equivalent critical behaviors.  Various {\em hard} and {\em
  soft} gauge-fixing implementations have been analyzed in noncompact
lattice Abelian gauge
models~\cite{KK-85,KK-86,BN-86,BN-86-b,BN-87,BPV-23}.  In this paper
we report a numerical study of the correlation functions of the gauge
and scalar fields, using a {\em hard} implementation of the Lorenz and
axial gauge fixing (i.e., implementing it as a constraint on the
lattice configurations), at the Coulomb-Higgs transition line where
the charged critical behavior emerges. Note that, in the hard Lorenz
gauge the scalar correlation functions are equivalent to the
gauge-invariant nonlocal correlators involving a smeared string of
gauge fields, which were introduced by Dirac
\cite{Dirac:1955uv,KK-85,KK-86}.  We also mention that the gauge
fixing can be considered as a particular breaking of the gauge
symmetry~\cite{Bonati:2021vvs}, which does not affect gauge-invariant
expectation values. Thus, this investigation may shed light on
continuous topological transitions in models with an emerging gauge
symmetry which is violated at the microscopic level.

The study of critical gauge-dependent correlations allows us to gain a
deeper understanding of the universality classes associated with
charged fixed points, where both gauge and matter fields develop a
critical behavior, which, for nongauge-invariant observables, somehow
depends on the specific form of the gauge fixing.  This situation
should be contrasted with what is observed along the transition lines
between the Coulomb and the molecular phase in the noncompact lattice
AH model, and at some continuous transitions of Abelian and
non-Abelian gauge theories with scalar matter, see, e.g.,
Refs.~\cite{BPV-21-ncAH, PV-19-AH3d, PV-19-CP, BPV-19-sqcd1,
  BPV-19-sqcd2, BPV-20-son, BFPV-21-sunfu, BPV-22-das}. In these cases
only matter fields play a role, while the gauge variables are
insensitive to the transition.

The paper is organized as follows. In Sec.~\ref{themodel} we define
the noncompact lattice AH model and summarize the main features of its
phase diagram for $N\ge 2$. In Sec.~\ref{gauobs} we define the gauge
fixings and the gauge-dependent observables which are studied
numerically.  In Sec.~\ref{FSsca} we report a numerical finite-size
scaling (FSS) study of these gauge-dependent correlations along the
continuous CH transition line.  Finally, in Sec.~\ref{conclu} we
summarize and draw our conclusions. In the Appendix, we report some
relations for the scalar correlation functions in soft and hard
gauges.

\section{Noncompact lattice AH model}
\label{themodel}

We consider the AH model defined on a cubic lattice of size $L^3$. The
fundamental variables are $N$-component complex scalar fields ${\bm
  z}_{\bm x}$ of unit length, i.e., satisfying $\bar{\bm z}_{\bm x}
\cdot {\bm z}_{\bm x} =1$, and noncompact real gauge variables
$A_{{\bm x},\mu}$.  Setting the lattice spacing $a=1$, the Hamiltonian
reads
\begin{equation}
\begin{aligned} \label{AHH}
H({\bm z},A) &= - J \, N
  \sum_{{\bm x},\mu} 2\,{\rm Re}\,( \bar{\bm z}_{\bm x} \cdot 
\lambda_{{\bm x},\mu}\,{\bm z}_{{\bm x}+\hat\mu})  \\ 
&+\frac{\kappa}{2} \sum_{{\bm x},\mu>\nu}
(\Delta_{\mu} A_{{\bm x},\nu} - \Delta_{\nu} A_{{\bm x},\mu})^2,
\end{aligned}
\end{equation}
where $\lambda_{{\bm x},\mu} \equiv e^{iA_{{\bm x},\mu}}$ and $\Delta_\mu
A_{{\bm x},\nu} = A_{{\bm x}+\hat{\mu},\nu}- A_{{\bm x},\nu}$.  The Hamiltonian
(\ref{AHH}) is invariant under the U($N$) transformations
\begin{equation}
{\bm z}_{\bm x} \to V {\bm z}_{\bm x},\qquad V\in\mathrm{U}(N),
\label{globaltra}
\end{equation}
and under the local gauge transformations
\begin{eqnarray}
{\bm z}_{\bm x} \to e^{i\Lambda_{\bm x}} {\bm
  z}_{\bm x}, \qquad
A_{{\bm x},\mu} \to A_{{\bm x},\mu} - \Delta_\mu \Lambda_{\bm x},
\label{gautra} 
\end{eqnarray}
where $\Lambda_{\bm x}$ is a real lattice function
and $\Delta_\mu \Lambda_{\bm x} = \Lambda_{{\bm x}+\hat\mu} - \Lambda_{\bm x}$ 
is a lattice derivative.

At variance with the compact formulation, the
partition function 
\begin{equation}
  Z = \int [d{\bm z}] [dA] \; e^{-H({\bm z},A)}
\label{Zpart}
\end{equation}
is only formally defined. Since the integration domain for the gauge variables
is noncompact, gauge invariance implies the divergence of $Z$ even on a finite
lattice.  The problem persists even when a maximal gauge fixing is added, if
periodic boundary conditions are chosen. Indeed, in this case the Hamiltonian
$H$ is also invariant under the group of noncompact transformations $A_{{\bm
x},\mu}\to A_{{\bm x},\mu} + 2\pi n_{\mu}$, where $n_{\mu}\in\mathbb{Z}$
depends on the direction $\mu$ but not on the point ${\bm x}$. This invariance
is also (at least partially) present in the gauge fixed theory, and therefore
$Z$ is ill-defined also when a gauge fixing is added.  To make the theory
well-defined on a finite lattice, we adopt $C^*$ boundary conditions
\cite{KW-91, LPRT-16, BPV-21-ncAH}, so that
\begin{equation}
A_{{\bm x} + L\hat{\nu},\mu}= -A_{{\bm x},\mu},\qquad
{\bm z}_{{\bm x}+L\hat{\nu}}= \bar{\bm z}_{{\bm x}}.
\label{cstarbc}
\end{equation}
$C^*$ boundary conditions preserve the local gauge invariance
(\ref{gautra}), provided one considers antiperiodic functions
$\Lambda_{\bm x}$. On the other hand, the U($N$) global symmetry
(\ref{globaltra}) is explicitly broken to O($N$) by the boundary
conditions.  This symmetry breaking is irrelevant for the bulk
behavior at continuous transitions, which would be still characterized
by the global U$(N)$ symmetry, since the critical behavior is
insensitive to the boundary conditions.

\begin{figure}[tbp]
\includegraphics*[width=0.85\columnwidth]{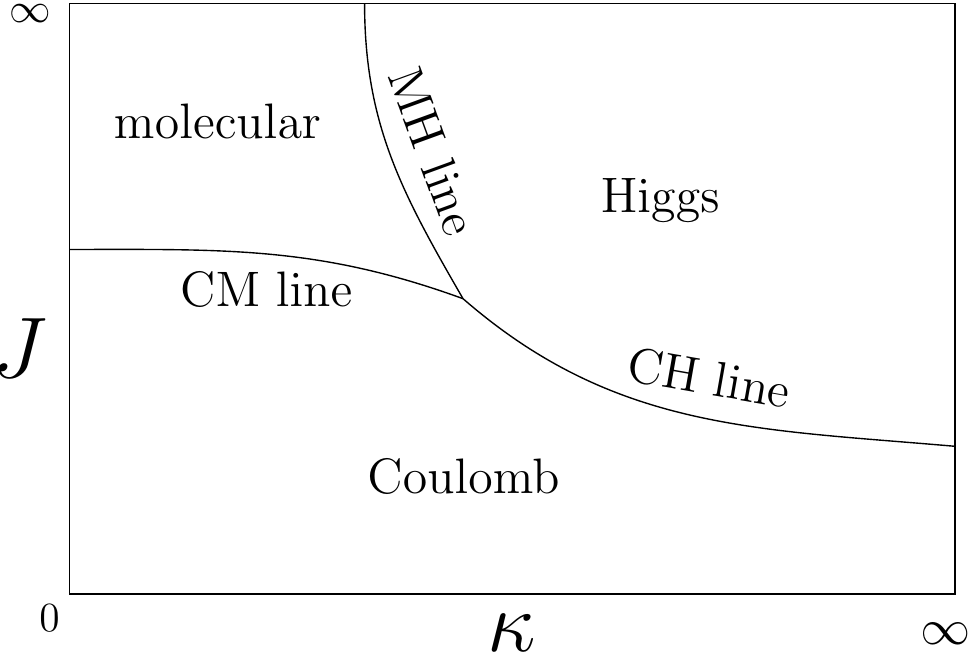}
  \caption{Sketch of the phase diagram of the lattice AH model with
    noncompact gauge fields and unit-length $N$-component complex
    scalar fields, for generic $N\ge 2$. There are three different
    phases, the Coulomb, Higgs and molecular phases, and three
    transition lines: the Coulomb-to-Higgs (CH) line between the
    Coulomb and Higgs phases, the Coulomb-to-molecular (CM) line, and
    the molecular-to-Higgs (MH) line. The model is equivalent to the
    CP$^{N-1}$ model for $\kappa=0$, to the O($2N$) vector model for
    $\kappa\to\infty$, and to the inverted $XY$ model for
    $J\to\infty$. }
\label{phdiasketchncLAH}
\end{figure}

The phase diagram of the noncompact multicomponent AH model is
reported in Fig.~\ref{phdiasketchncLAH} (see e.g.
Refs.~\cite{BPV-21-ncAH,BPV-23} for more details). For any $N\ge 2$
the phase diagram presents three phases. There is a Coulomb phase for
small $J$, in which the global SU($N$) symmetry is unbroken and the
correlations of the gauge field are long ranged. There are two
different phases for large $J$, where the global SU($N$) symmetry is
spontaneously broken. They are distinguished by the behavior of the
gauge modes. In the Higgs phase (large $\kappa$), the correlations of
the gauge field are gapped, while in the molecular phase (small
$\kappa$) the gauge field is ungapped.  The Coulomb, molecular, and
Higgs phases are separated by various transition lines, see
Fig.~\ref{phdiasketchncLAH}: the CM line between the Coulomb and the
molecular phase, the MH line between the molecular and the Higgs
phases, and the CH line between the Coulomb and the Higgs phase. The
transitions occurring along these lines may be of first order or
continuous, depending on the number $N$ of
components~\cite{BPV-21-ncAH}.  In particular, the transitions along
the CH line are continuous for $N > N^\star$, with~\cite{BPV-21-ncAH}
$N^{\star}=7(2)$, and of first order in the opposite case.  For $N >
N^\star$, the critical behavior is controlled by the stable charged
fixed point of the AH field theory.  In the following we focus on the
critical behavior along the CH line for $N=25$, focusing on the
critical behaviors of the gauge-dependent correlation functions of the
gauge and scalar fields in the presence of gauge fixings.

\section{Gauge fixing and observables}
\label{gauobs}

\subsection{Lorentz and axial gauge fixings}
\label{gaugefixing}

To study the properties of the correlation functions of
gauge-dependent operators, such as the gauge field $A_{{\bm x},\mu}$
and the scalar field ${\bm z}_{\bm x}$, a gauge fixing is
required. For this purpose, one may add a {\em hard} constraint in the
partition function, i.e.,
\begin{equation}
  Z_{\rm gf}= \int [d{\bm z}] [dA] \; \prod_{\bm x} 
  \delta[F_{\bm x}(A)]\, e^{-H({\bm z},A)},
\label{Zhard}
\end{equation}
where $F_{\bm x}(A)$ is a local linear function of the gauge
variables.  Here, we consider the hard Lorenz gauge fixing,
corresponding to
\begin{equation}
  F_{L,{\bm x}}(A) = \sum_{\mu} \Delta^{-}_\mu A_{{\bm x},\mu}
   = \sum_{\mu}  (A_{{\bm x},\mu}-A_{{\bm x}-\hat{\mu},\mu})=0,
\label{Lorenzgau}
\end{equation} 
and the axial gauge fixing with
\begin{equation}
F_{A,{\bm x}}(A) = A_{{\bm x},3}=0.
\label{axialgau}
\end{equation}
In perturbation theory one often relies on soft gauge fixings,
obtained by adding a term
\begin{equation}
H_{\rm gf} = {1\over 2\zeta} G(A), \qquad G(A) = \sum_{\bm x} F_{{\bm x}}(A)^2,
\label{H-soft}
\end{equation}
to the Hamiltonian.  The hard gauge fixing is obtained by taking the
limit $\zeta \to 0$. The behavior of nongauge-invariant correlation
functions for $\zeta\not=0$ can be related to that of the same
quantities computed in the hard case $\zeta=0$, see the Appendix for
scalar-field correlations and Ref.~\cite{BPV-23} for gauge-field
correlations. In this work we mainly consider hard gauge
fixings. Some results for the soft Lorenz case---the one which is used
in perturbation theory---are reported in Sec.~\ref{scacosoft}.

\subsection{Observables}
\label{obds}

Numerical studies of the lattice AH model with multicomponent scalar
fields have mostly focused on gauge-invariant observables,
such as the correlations of the scalar bilinear operator
\begin{equation}\label{eq:Q}
Q_{{\bm x}}^{ab} = \bar{z}_{\bm x}^a z_{\bm x}^b -
\frac{1}{N} \delta^{ab},
\end{equation}
which is gauge invariant and satisfies $Q_{\bm x}^\dagger=Q_{\bm x}$.
Note that $Q_{{\bm x}}^{ab}$ is not irreducible under the O($N$) subgroup
of transformations that is not broken by the $C^*$ boundary conditions,
but it is the sum 
of $Q_S = (Q + Q^T)/2$ and $Q_A = (Q - Q^T)/2$ (where $Q^T$ is the 
transpose of $Q$) that instead transform irreducibly.

To define the two-point correlation function of $Q$, we set
\begin{equation}
   \widetilde{Q}^{ab}({\bm p}) = \sum_{\bm x} e^{i{\bm p}\cdot {\bm x}} 
     Q_{\bm x}^{ab}.
\end{equation}
Then, we define
\begin{equation}
\label{FTG}
\begin{aligned}
\widetilde{G}_Q({\bm p}) &= 
   {1\over L^3} \langle {\rm Tr}\,  \widetilde{Q}(-{\bm p})
   \widetilde{Q}({\bm p}) \rangle  \\
   &=
   {1\over L^3	} \sum_{{\bm x}{\bm y}}
     e^{i {\bm p}\cdot ({\bm x} - {\bm y}) } 
     G_Q({\bm x}-{\bm y}), 
\end{aligned}
\end{equation}
where ${\bm x}$ and ${\bm y}$ are the sites of the lattice,
and~\footnote{In the presence of $C^*$ boundary conditions,
$Q_{{\bm x}+L\hat{\nu}}=Q_{\bm x}^T$, so that 
$G_Q({\bm x} + L\hat{\nu})$
is not directly related to $G_Q({\bm x})$. As a consequence, 
there are some subtleties in the definition of the Fourier transform.
Working for simplicity in one dimension and assuming 
that $0\le x, y \le L-1$, Eq.~(\ref{gxyp})
defines $G_Q(x)$ for $-L+1\le x\le L-1$, with $G_Q(x) = G_Q(-x)$
($0\le x \le L-1$). Since $G_Q(x) \not= G_Q(x +L)$, we also obtain 
$G_Q(x) \not=  G_Q(L-x)$. As a consequence, $\widetilde{G}_Q(p)$
defined in Eq.~(\ref{FTG}) differs from 
$\widehat{G}_Q(p) = \sum_{x=0}^{L-1} e^{ipx} G_Q(x)$, 
with $G_Q(x)$ defined in Eq.~(\ref{gxyp}). For instance, for $p=0$, 
the two quantities are related as $\widetilde{G}_Q(0) = 
\widehat{G}_Q(0) - (1/L) \sum_{x=1}^{L-1} x [G_Q(x) - G_Q(L-x)]$.
}
\begin{equation}
G_Q({\bm x}-{\bm y}) = \langle {\rm Tr}\, Q_{\bm x} Q_{\bm y} \rangle.
\label{gxyp}
\end{equation}
We define a susceptibility $\chi_Q$ and a correlation length $\xi_Q$
as
\begin{eqnarray}
  \chi_Q &=& \widetilde{G}_Q({\bm 0}), \\ 
 \xi_Q^2 &=& {1\over 4 \sin^2 (\pi/L)} {\widetilde{G}_Q({\bm 0}) -
   \widetilde{G}_Q({\bm p}_m)\over \widetilde{G}_Q({\bm p}_m)},\quad
\label{xidefpb}
\end{eqnarray}
where ${\bm p}_m = (2\pi/L,0,0)$. A particulary useful quantity is the
RG invariant ratio
\begin{equation}
  R_Q \equiv \xi_Q/L.
\label{rxidef}
\end{equation}
As already anticipated, in this paper we extend our analysis to
correlations of gauge-dependent operators, such as the gauge field
$A_{{\bm x},\mu}$ and the scalar field ${\bm z}_{\bm x}$.

For the gauge field, we set
\begin{equation}
  \widetilde{A}_\mu({\bm p}) = e^{i p_\mu/2} \sum_{\bm x} e^{i{\bm p}\cdot{\bm x}}
  A_{{\bm x},\mu},
  \label{tildeA}
  \end{equation}
where the prefactor takes into account that the gauge field is defined
on a link (it guarantees that $\widetilde{A}_\mu({\bm p})$ is odd
under reflections in momentum space). Then, we define the correlation function
\begin{eqnarray}
  \widetilde{C}_{\mu\nu}({\bm p}) = L^{-3} \, \langle
  \widetilde{A}_\mu({\bm p}) \widetilde{A}_\nu(-{\bm p})\rangle.
  \label{twopA}
  \end{eqnarray}
Due to the $C^*$ boundary conditions, $A_{{\bm x},\mu}$ is
antiperiodic, so that the allowed momenta ${\bm p}$ of its lattice
Fourier transform are $p_i = \pi (2 n_i + 1)/L$ with $ n_i = 0,\ldots
L-1$ (in particular, ${\bm p} = 0$ is not allowed).  Note that the
charge-conjugation symmetry, i.e., the symmetry under the
transformations $A_{{\bm x},\mu}\to -A_{{\bm x},\mu}$ and ${\bm
  z}_{\bm x}\to\bar{\bm z}_{\bm x}$, is preserved by the $C^*$
boundary conditions and by the Lorenz and axial gauge fixings, so that
we have $\langle A_{{\bm x},\mu}\rangle=0$.

We define a gauge-field susceptibility as (no sum on repeated indices implied)
\begin{eqnarray}
  \chi_{A} = \widetilde{C}_{\mu\mu}({\bm p}_a),
   \label{chiAdef}
\end{eqnarray}
where $\mu$ is one of the directions and ${\bm p}_a$ is one of the smallest
momenta compatible with the antiperiodic boundary conditions of $A_{{\bm
x},\mu}$, i.e., 
\begin{equation}
{\bm p}_a=(\pi/L, \pi/L, \pi/L).
  \label{padef}
\end{equation}
In the case of the Lorenz gauge, all components are equivalent. In the
axial case in which $A_{{\bm x},3} = 0$, we take either $\mu=1$ or $\mu=2$.
The second-moment correlation length of the gauge field is defined by
\begin{equation}\label{xiA}
\xi^2_{A} = 
\frac{1}{(\hat{p}_a^2-\hat{p}_b^2) }
\frac{\widetilde{C}_{\mu\mu}({\bm p}_b) - \widetilde{C}_{\mu\mu}({\bm p}_a)}
{\tilde{C}_{\mu\mu}({\bm p}_a)}\,,
\end{equation}
where  
\begin{equation}
  \hat{p}^2=\sum_{\mu=1}^3 4\sin^2(p_\mu/2),
  \quad
  {\bm p}_b={\bm p}_a+{2\pi \over L} \hat{\nu},
\end{equation}
where, somewhat arbitrarily, we have taken $\nu\not=\mu$. 
In the Lorenz case any pair of directions $\mu,\nu$ is equivalent.
In the axial case we fix $\nu = 3$ and $\mu=1$ or 2, as $A_{{\bm x},3}=0$.
The Binder cumulant of the gauge field is defined by
\begin{eqnarray}
  U_{A} = \frac{\langle m_{2,\mu}^2\rangle}{\langle m_{2,\mu} \rangle^2},
  \qquad
m_{2,\mu} = \big|\sum_{\bm x} e^{i{\bm p}_a\cdot {\bm x}} A_{{\bm x},\mu}\big|^2.
    \label{binderdef2}
\end{eqnarray}
with $\mu\not=3$ in the axial case.

Gauge field correlations in the Lorenz and axial gauges are strictly related
\cite{BPV-23}.
In particular, we have 
\begin{eqnarray}
  \chi_{A}^{(A)} = 3 \chi_{A}^{(L)}, \qquad 
  U_{A}^{(A)} = U_{A}^{(L)}. 
  \label{chiconn}
\end{eqnarray}
where the superscripts $(A)$ and $(L)$ denote the quantities computed
in the axial and Lorenz gauge, respectively (we will consistently use
this notation, whenever needed).

Finally, we study the critical behavior of the correlation function of
the scalar field. Again, note that the scalar variable ${\bm z}_{\bm
  x}$ is not periodic, so that one should be careful in the definition
of the Fourier-transformed two-point function. We set
\begin{equation}
  \widetilde{\bm z}({\bm p}) = \sum_{\bm x}
  e^{i{\bm p}\cdot {\bm z}} {\bm z}_{\bm x}, \qquad 
\widetilde{G}_z({\bm p}) = {1\over L^3}
    \left\langle \left| \widetilde{\bm z}({\bm p}) \right|^2 \right\rangle.
\label{gxypz}
\end{equation}
We define the corresponding susceptibility $\chi_z$ and length scale
$\xi_z$ as 
\begin{eqnarray}
\chi_z = \widetilde{G}_z({\bm 0}),
\quad
\xi_z^2 \equiv {1\over 4 \sin^2 (\pi/L)} {\widetilde{G}_z({\bm 0}) -
   \widetilde{G}_z({\bm p}_m)\over \widetilde{G}_z({\bm p}_m)},\quad
\label{xidefpbz}
\end{eqnarray}
with ${\bm p}_m = (2\pi/L,0,0)$. 
Finally, we define 
the Binder cumulant for the scalar field by
\begin{eqnarray}
  U_{z} = 
\frac{\langle m_{2}^2\rangle}{\langle m_{2} \rangle^2}, \qquad
m_{2} = \sum_{{\bm x},{\bm y}} \bar{\bm z}_{\bm x} \cdot {\bm z}_{\bm y}.
    \label{binderdefz}
    \end{eqnarray}

It is interesting to observe that the analysis of the gauge-dependent
correlation functions of the gauge and scalar field 
also provides information on the behavior of gauge-invariant quantities.
To characterize the gauge-field behavior, 
one might also consider the field-strength 
gauge-invariant correlation function
\begin{equation}
G_{F}(\bm{x} - \bm{y}) = \sum_{\mu\nu} 
   \langle F_{{\bm x},\mu\nu} F_{{\bm y},\mu\nu} \rangle
\end{equation}
with $F_{{\bm x},\mu\nu} = \Delta_\mu A_{{\bm x},\nu} - 
\Delta_\nu A_{{\bm x},\mu}$. This quantity is trivially related to the 
correlation function of the gauge field $A_{{\bm x},\mu}$. This relation 
is particulary simple in the hard Lorenz gauge. In Fourier space  we have 
\begin{equation}
\sum_{\mu\nu} \langle \widetilde{F}_{\mu\nu}(-{\bm p}) 
   \widetilde{F}_{\mu\nu} ({\bm p}) \rangle = 
2 \hat{p}^2 \sum_\nu \langle \widetilde{A}_{\nu}(-{\bm p})
   \widetilde{A}_{\nu} ({\bm p}) \rangle.
\end{equation}
Although mathematically equivalent, it is clear that the analysis 
of the gauge susceptibility $\chi_A$ defined in Eq.~(\ref{chiAdef})
  is much more convenient, from a numerical point of view, 
than the analysis of the 
field-strength susceptibility, which does not diverge, but rather vanishes 
in the infinite-volume limit $L\to \infty$.

The scalar-field correlator can also be related to  a gauge-invariant 
correlation function, although only in the presence of a hard Lorenz
gauge fixing.  Indeed, let $V(\bm{x},{\bm y})$ be the 
lattice Coulomb potential due to a unit charge 
in ${\bm y}$, i.e., the solution of the lattice equation 
(the lattice derivatives act on ${\bm x}$)
\begin{equation}
  \sum_\mu \Delta^-_\mu \Delta_\mu V({\bm x},{\bm y}) = -
   \delta_{{\bm x},{\bm y}}.
\end{equation}
Then, we can define a dressed nonlocal scalar
operator~\cite{Dirac:1955uv,KK-85,KK-86}
\begin{eqnarray}
&&  {\bm \Gamma}_{\bm x} = 
  \exp[i \sum_{{\bm y},\mu} E_\mu({\bm y},{\bm x}) 
    A_{{\bm y},\mu}] \, {\bm z}_{\bm x}, \label{phixdef}\\
&&E_{\mu}({\bm y},{\bm x}) = V({\bm y}+\hat{\mu},{\bm x}) -
V({\bm y},{\bm x}).
\end{eqnarray}
Such extended operator ${\bm \Gamma}_{\bm x}$ is gauge invariant and
satisfies ${\bm \Gamma}_{\bm x} = {\bm z}_{\bm x}$ in the hard Lorenz
gauge.  Therefore, the scalar-field correlators within the hard Lorenz
gauge are equivalent to the gauge-invariant correlators of the
extended field ${\bm \Gamma}_{\bm x}$.

\section{Critical behavior of the gauge-dependent observables}
\label{FSsca}

In this section we will discuss the critical behavior of
gauge-dependent observables, focusing on the CH transition line in
Fig.~\ref{phdiasketchncLAH} for a system with $N=25$ components.  As
extensively discussed in Refs.~\cite{BPV-21-ncAH, Bonati:2021vvs,
  BPV-22}, for this value of $N$, the transitions along the CH line
are associated with the stable RG fixed point of the continuum
AH field theory (\ref{AHFT}).  In particular, we have
\cite{BPV-22} $\nu= 0.817(7)$ and $\eta_Q = 0.882(2)$ for $N=25$
($\eta_Q$ is the susceptibility exponent for $\chi_Q$ defined in
Eq.~(\ref{xidefpb})), in substantial agreement with the large-$N$
estimates~\cite{HLM-74, DHMNP-81, YKK-96, MZ-03, KS-08, BPV-22}.

To determine the behavior along the CH line, we fix $\kappa= 0.4$ and
vary $J$ close to the critical point \cite{Bonati:2021vvs} $J=J_c =
0.295515(4)$.  The numerical setup used for the simulations is the
same as in Ref.~\cite{BPV-23}, to which we refer for more details.

\subsection{Finite-size scaling}
\label{fssobs}

We summarize here the main FSS equations that we exploit in our
numerical analysis.  We consider RG invariant quantities, such as the
ratios $R_Q=\xi_Q/L$, $\xi_{A}/L$, $\xi_z/L$ and the Binder parameters
$U_Q$, $U_{A}$, $U_z$, which scale, in the large-$L$ limit, as
\begin{equation}
\label{eq:FSS1}
R(J,L)={\cal R}(X)+ O(L^{-\omega}),\quad X=(J-J_c)L^{1/\nu},
\end{equation}
where ${\cal R}$ is a universal function apart from a normalization of
the argument $X$, $J_c$ is the critical value, $\nu$ is the critical
correlation-length exponent, and $\omega$ is the exponent controlling
the leading scaling corrections~\cite{PV-02}.  Ref.~\cite{BPV-22}
estimated $\nu=0.817(7)$ and $\omega \approx 1$ for the continuous CH
transition with $N=25$.

If one of the RG-invariant quantities is monotonic with respect to
$J$---this is the case for the ratio $R_Q\equiv
\xi_Q/L$~\cite{BPV-21-ncAH,BPV-22}---the argument of the scaling
function ${\cal R}(X)$ of a different RG-invariant quantity $R$ can be
replaced by $R_Q$, formally writing $X={\cal
  R}_Q^{-1}(R_Q)$. Therefore, we can write the asymptotic FSS behavior
as
\begin{equation}
\label{eq:FSS2}
R(J,L)=\widehat{\cal R}(R_Q) + O(L^{-\omega}),
\end{equation}
where $\widehat{\cal R}$ is a universal function as well, which is
now completely determined, with no further need of fixing 
nonuniversal normalizations. The FSS relation
(\ref{eq:FSS2}) is particuraly convenient because it allows us to
check universality between different models in a completely unbiased
way, without requiring any parameter tuning.

The RG dimension $y_o$ of a local operator $O_{\bm x}$ can be estimated 
by analyzing the corresponding susceptibility $\chi_o$ defined in terms of 
the Fourier-transformed two-point correlation function computed in the 
small-momentum limit. In the FSS limit, it behaves as
\begin{equation}
  \chi_{o}\approx L^{d-2 y_o} F_{\chi}(R_Q) = L^{2-\eta_o}
  F_{\chi}(R_Q),
\label{chiosca}
\end{equation}
where $F_\chi$ is a universal function apart from a nonuniversal
multiplicative factor, and we used the standard RG relation
$y_{o}=(d-2+\eta_{o})/2$.  Note that these relations hold for
$y_{o} < d/2$, or for $\eta_o<2$. Otherwise, the analytic background 
is the dominant contribution and the nonanalytic scaling part 
represents a correction term.

\subsection{The gauge-field correlations}
\label{gaufico}

\subsubsection{RG-invariant quantities}
\label{RGinv}

\begin{figure}
\includegraphics*[width=0.95\columnwidth]{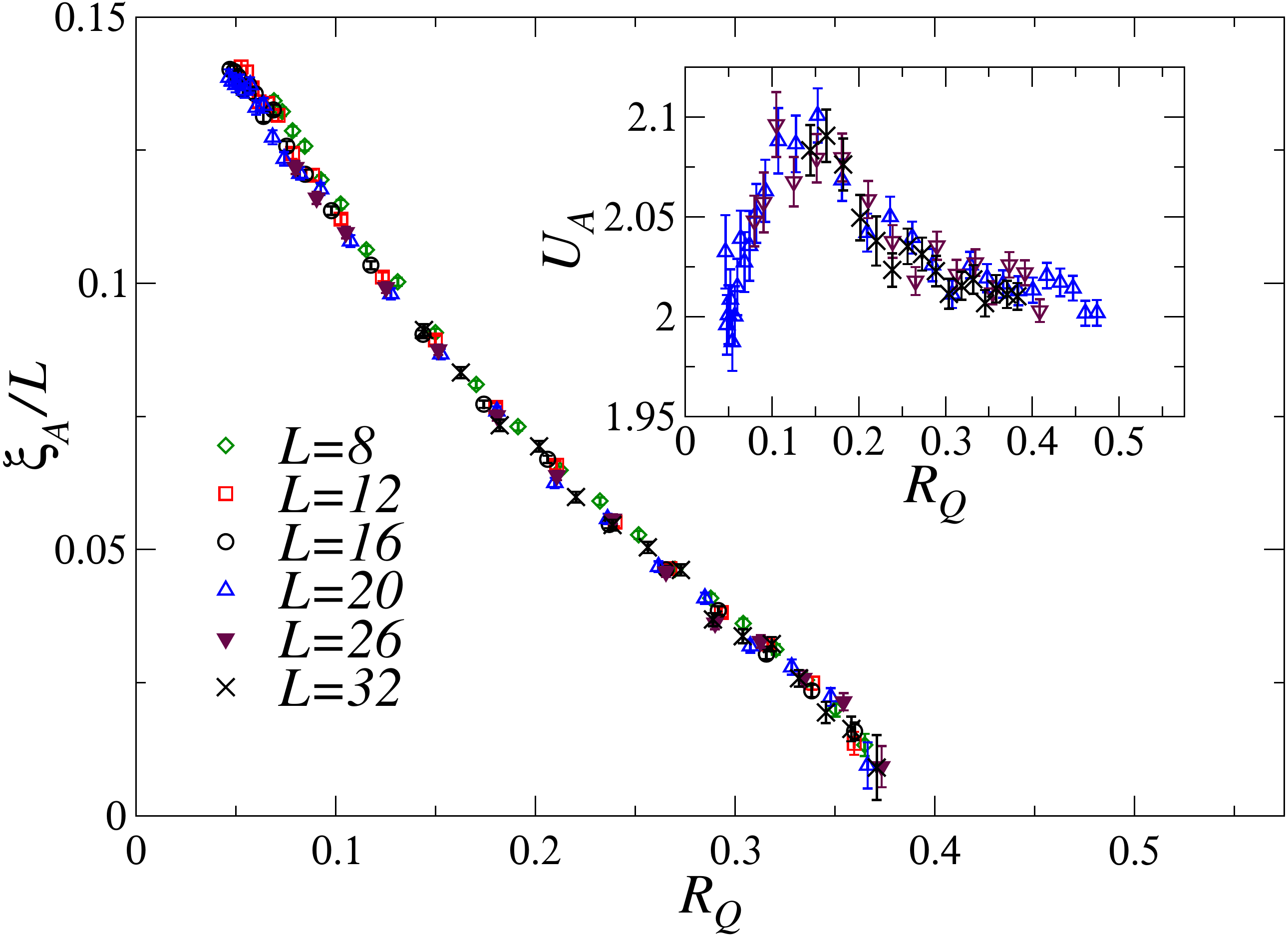}
\caption{Plot of the ratio $\xi_A/L$ and of $U_A$ (inset) as 
  a function of $R_Q$, in the Lorenz gauge.
  For the sake of readability, in the inset only the estimates of 
   $U_A$ for the three larger lattices are reported.
}
\label{critALorenz}
\end{figure}

\begin{figure}
\includegraphics*[width=0.95\columnwidth]{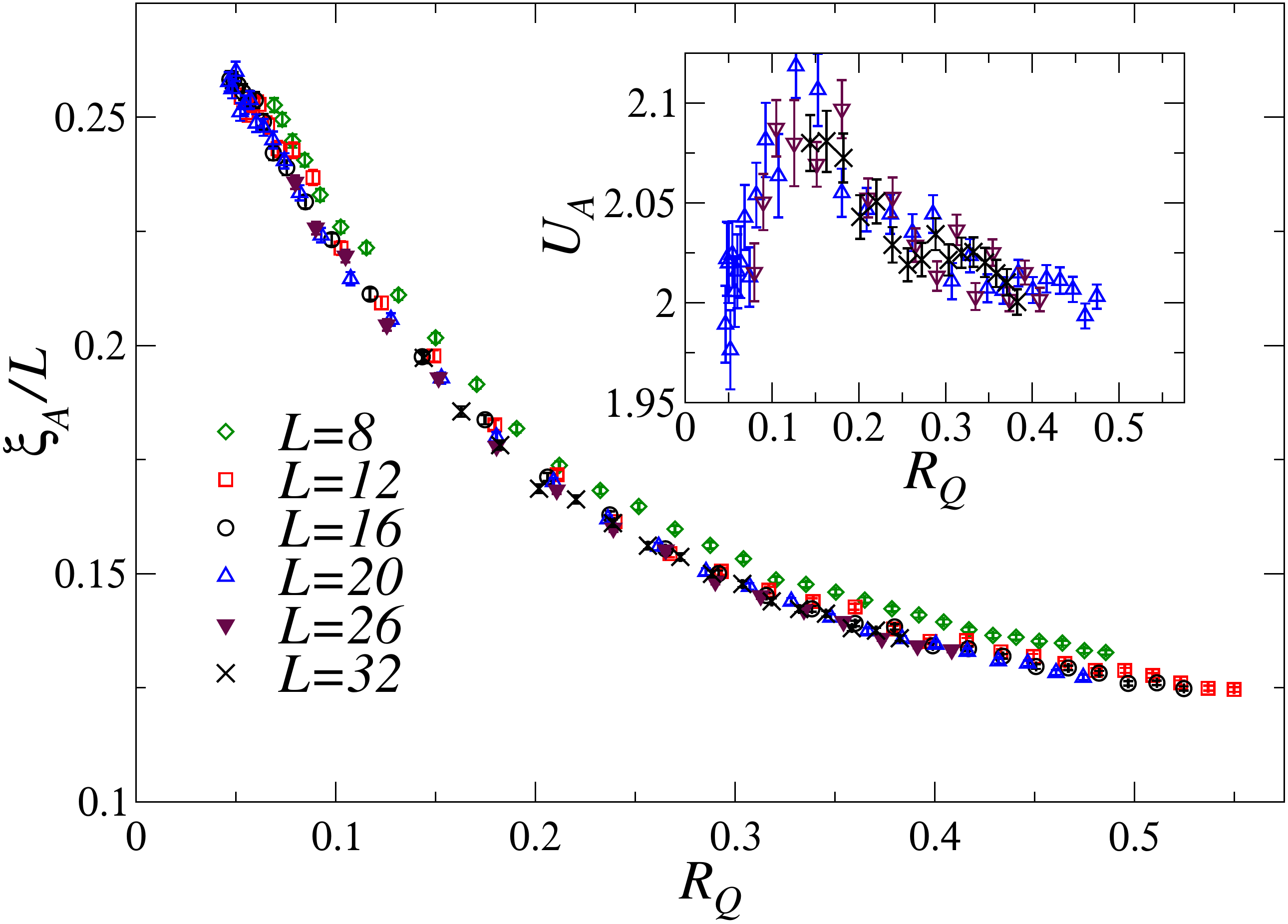}
\caption{Plot of the ratio $\xi_A/L$ and of $U_A$ (inset) as a
  function of $R_Q$, in the axial gauge.  For the sake of readability,
  in the inset only the estimates of $U_A$ for the three larger
  lattices are reported.  }
\label{critAaxial}
\end{figure}

To verify that gauge-field correlations are critical at the
transition, in Figs.~\ref{critALorenz} and \ref{critAaxial} we report
the correlation length $\xi_A$ and the Binder cumulant $U_A$ for the
Lorenz and axial gauge fixing, respectively. Data clearly scale
according to Eq.~(\ref{eq:FSS2}) in both cases. Moreover, the data for
$U_A$ are consistent with the relation (\ref{chiconn}). The behavior
of $U_A$ for $R_Q \to 0$ and $\infty$ can be easily determined. For
$R_Q\to 0$, the Binder parameter should converge to the value it
assumes in the Coulomb phase. An easy computation for $J=0$ gives $U_A
= 2$. For $R_Q\to \infty$ it should converge to the value in the Higgs
phase, where $A_{{\bm x},\mu}$ is disordered, thus one easily obtain
$U_A = 2$. Note that it is crucial here that the Binder parameter is
computed at a nonzero value of the momentum. Indeed, for a
one-component real scalar field, the Binder parameter computed at zero
momentum (this would be the case if periodic boundary conditions were
used) would converge to 3 in the disordered phase.  The numerical
results are consistent with these predictions, see
Figs.~\ref{critALorenz} and \ref{critAaxial}. The FSS curves for
$\xi_A/L$ in the two gauges differ, although they appear qualitatively
similar.  We mention that $\xi_A$ is quite noisy for the Lorenz gauge,
and therefore it could be measured with a reasonable accuracy only for
$R_Q\lesssim 0.37$.

\subsubsection{Scaling behavior of the susceptibilities}
\label{etaAest}

\begin{figure}
\includegraphics*[width=0.95\columnwidth]{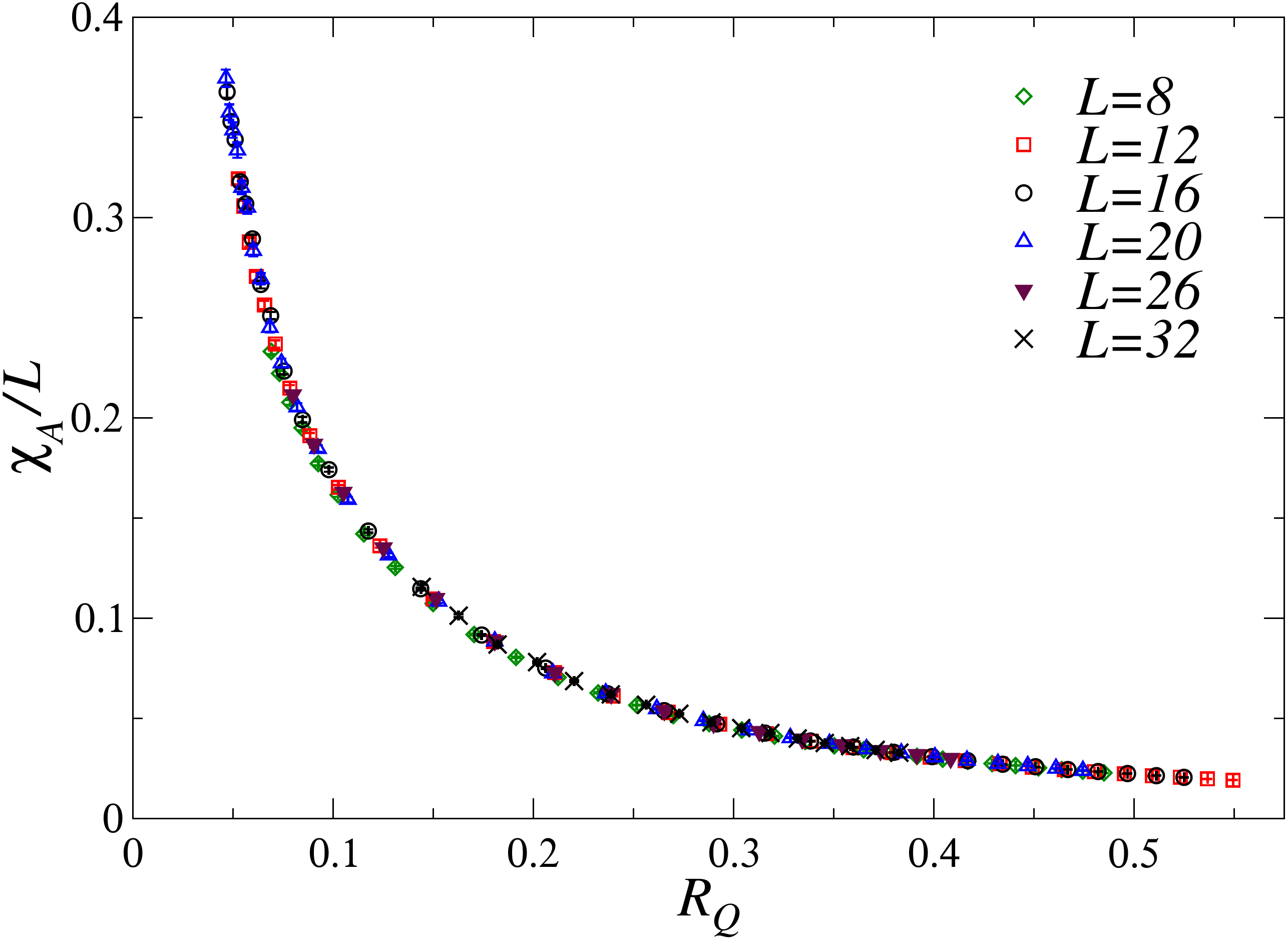}
\caption{Plot of $\chi_A/L^{2-\eta_A} = \chi_A/L$ for $\eta_A = 1$ 
as a function of $R_Q$ in the Lorenz gauge. 
}
\label{fig:chiA}
\end{figure}

Let us now consider the susceptibility $\chi_A$ of the gauge field
that is expected to scale as in Eq.~(\ref{chiosca}) with $\eta_A = 1$
\cite{ZJ-book,HT-96,BPV-23} for both gauge fixings, independently of
$N$. This prediction is nicely confirmed by the results shown in
Fig.~\ref{fig:chiA} for the Lorenz gauge. Axial-gauge results are
consistent.  To obtain an unbiased estimate of $\eta_A$ from the data,
we have performed a simultaneous analysis of the axial and Lorenz
results, using the exact relation~\eqref{chiconn}. Data have first
been fitted to the expected FSS behavior (\ref{chiosca}), using a
polynomial approximation for the function $F_\chi(R_Q)$, and including
each time only data with $L \ge L_{\rm min}$. We find estimates of
$\eta_A$ that increase with $L_{\rm min}$, varying from $\eta_A =
0.981(1)$ to $0.993(3)$ ($L_{\rm min} = 8$ and 20, respectively). We
have then included scaling corrections.  Fixing $\omega = 1$, as
already done in Ref.~\cite{BPV-22}, we obtain stable results,
providing the estimate
\begin{equation}
  \eta_A = 1.004(5).
  \label{etaest}
  \end{equation}
Numerical data, therefore, are perfectly consistent with the
field-theoretical prediction $\eta_A = 1$.

Finally, we have analyzed the susceptibility $\chi_B$ of the photon-mass 
operator $B_{\bm x} = \sum_\mu A_{{\bm x},\mu}^2$,
\begin{equation}
\chi_B = {1\over V} \sum_{{\bm x}{\bm y}} 
   \left\langle (B_{\bm x} - \langle B\rangle) 
                (B_{\bm y} - \langle B\rangle)  \right\rangle.
\end{equation}
Since the photon mass renormalizes as the gauge field in the AH
field theory with Lorenz gauge fixing~\cite{ZJ-book}, we expect 
the RG dimension $y_B$ of $B_{\bm x}$ to be  given by
\begin{equation}
  y_B = 1 + \eta_A = 2,
  \label{ybft}
  \end{equation}
which implies $\eta_B = 2 y_B - 1 = 1 + 2 \eta_A = 3$.  Note that,
since $\eta_B > 2$, the FSS equation (\ref{chiosca}) does not provide
the leading asymptoptic behavior, being suppressed with respect to the
$O(1)$ contributions from the analytic background~\cite{PV-02}. In
this case, at the critical point we expect that $\chi_B(J_c) = a + b
L^{2-\eta_B}$, where $a$ is the contribution of the analytic
background.  We have performed fits of $\chi_B(J_c)$ to $a + b
L^{2-\eta_B}$, including only data satisfying $L\ge L_{\rm min}$.  We
obtain $\eta_B = 2.34(5)$, 2.6(1), 3.0(3), for $L_{\rm min} =
8,12,16$. Results are reasonably consistent with the field-theory
prediction.

\subsection{The scalar-field correlations}
\label{scalarcrit}

\begin{figure}
\includegraphics*[width=0.95\columnwidth]{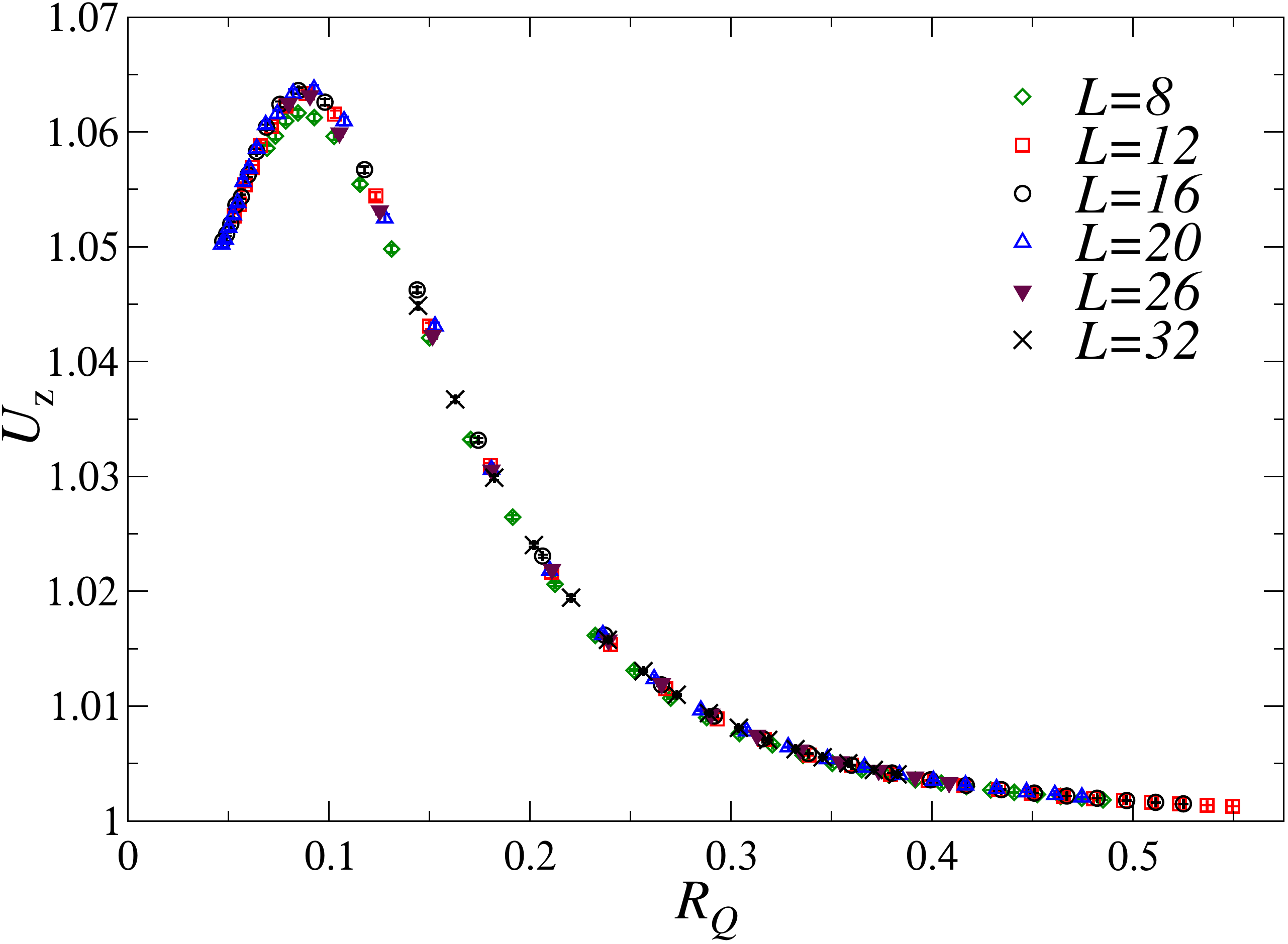}
\includegraphics*[width=0.95\columnwidth]{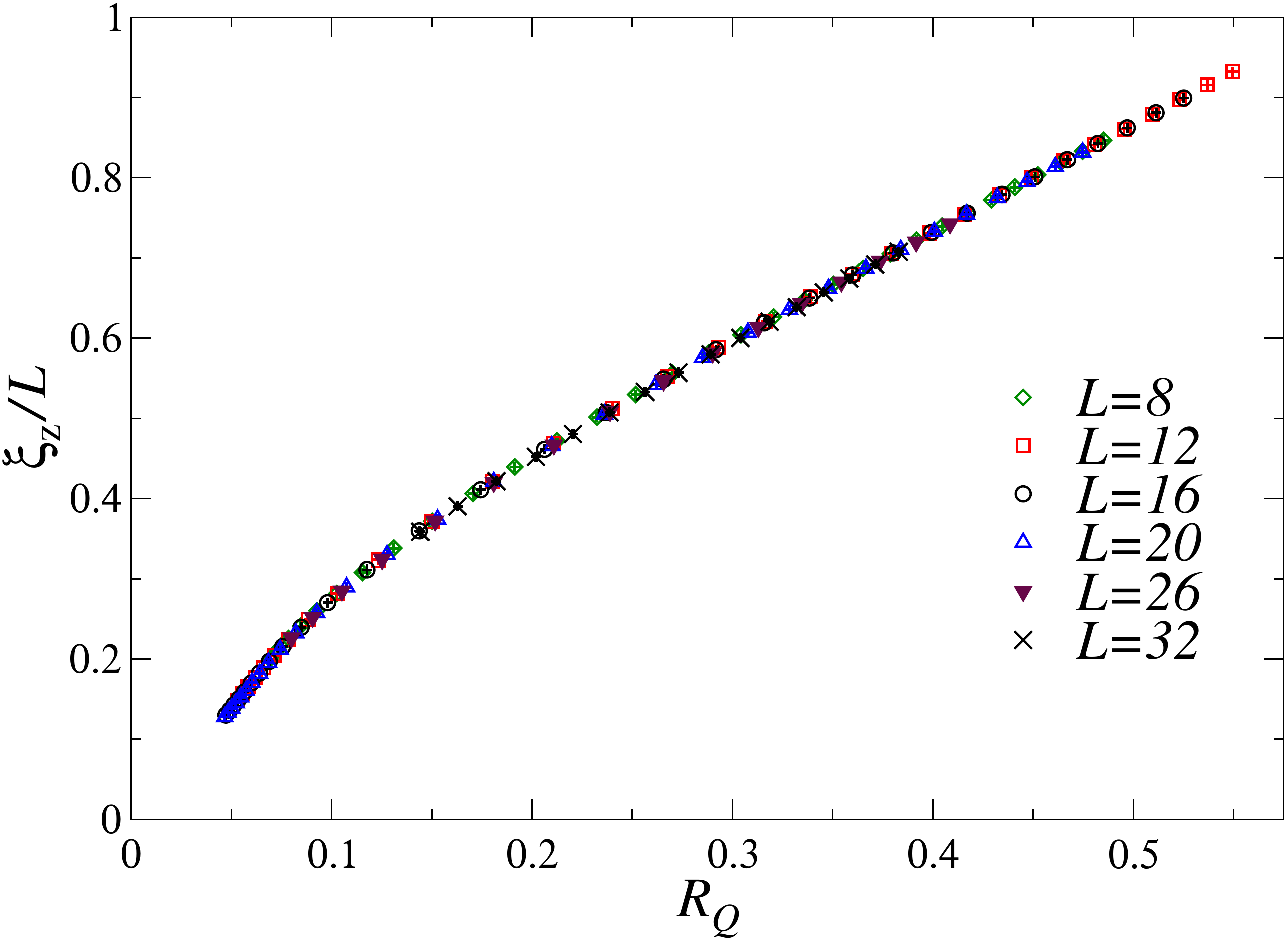}
\caption{Plot of the ratio $\xi_z/L$ (bottom) and $U_z$ (top) defined
  in terms of the gauge-dependent correlation function of the scalar
  field ${\bm z}_{\bm x}$, see Eq.~(\ref{gxypz}), for the hard Lorenz
  gauge.  For $R_Q\to 0$, we have $U_z = (N+1)/N = 1.04$, while
  $U_z=1$ for $R_Q\to\infty$.}
\label{critzLorenz}
\end{figure}

Let us now discuss the behavior of scalar-field correlations at the
transition. In Fig.~\ref{critzLorenz} we report the Binder cumulant
$U_z$ and the correlation ratio $\xi_z/L$ versus $R_Q$ for the model
with a hard Lorenz gauge fixing. Data are consistent with the FSS
relation (\ref{eq:FSS2}), with very small scaling corrections,
indicating that, in this gauge, scalar fields develop long-range
correlations at the transition.  Note that $\xi_z/L$ increases sharply
and $U_z\to 1$ as $R_Q$ increases, so that we can conclude that also
in the Higgs phase the scalar-field two-point function (\ref{gxypz})
is long ranged.  Note that, for the one-component model, it was
rigorously proved that the scalar field condenses in some region of
the Higgs phase~\cite{KK-85,KK-86,BN-87}. Our results suggest that
this property extend to multicomponent AH models and characterizes the
whole Higgs phase.

\begin{figure}
  \includegraphics*[width=0.95\columnwidth]{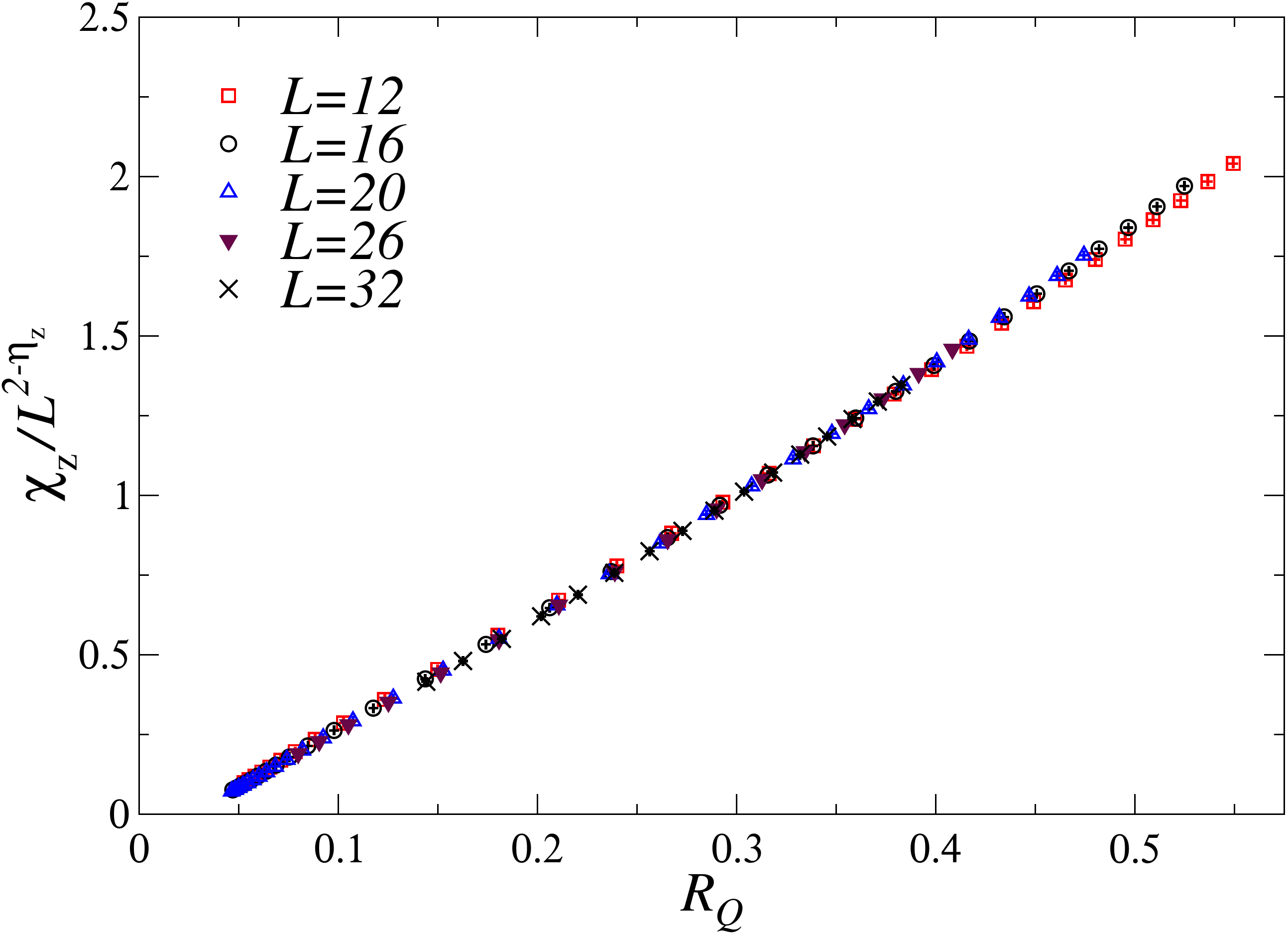}
  \caption{Plot of $\chi_z L^{-2 + \eta_z}$ as a function of $R_Q$,
    for the hard Lorenz gauge. We use the numerical estimate
    (\ref{yzdef}), i.e. $\eta_z = -0.069$.}
\label{chizscal}
\end{figure}

We have also determined the exponent $\eta_z$ of the scalar
susceptibility.  We have performed fits to Eq.~(\ref{chiosca}) and
also fits that take scaling corrections into account with $\omega
\approx 1$. The results are fully consistent and give
\begin{equation}
\eta_z = - 0.069(1),  
  \label{yzdef}
  \end{equation}
which allows us to determine the RG dimension of the scalar field:
$y_z = (1 + \eta_z)/2 = 0.4655(5)$. The quality of the scaling, see
Fig.~\ref{chizscal}, is excellent. It is interesting to observe that
the estimate (\ref{yzdef}) is close to the large-$N$
expansion~\cite{HLM-74,KS-08}
\begin{equation}
  \eta_z = - {20\over \pi^2} {1\over N} + O(N^{-2}),
\end{equation}
that predicts $\eta_z \approx -0.081$ for $N=25$.

\begin{figure}
  \includegraphics*[width=0.95\columnwidth]{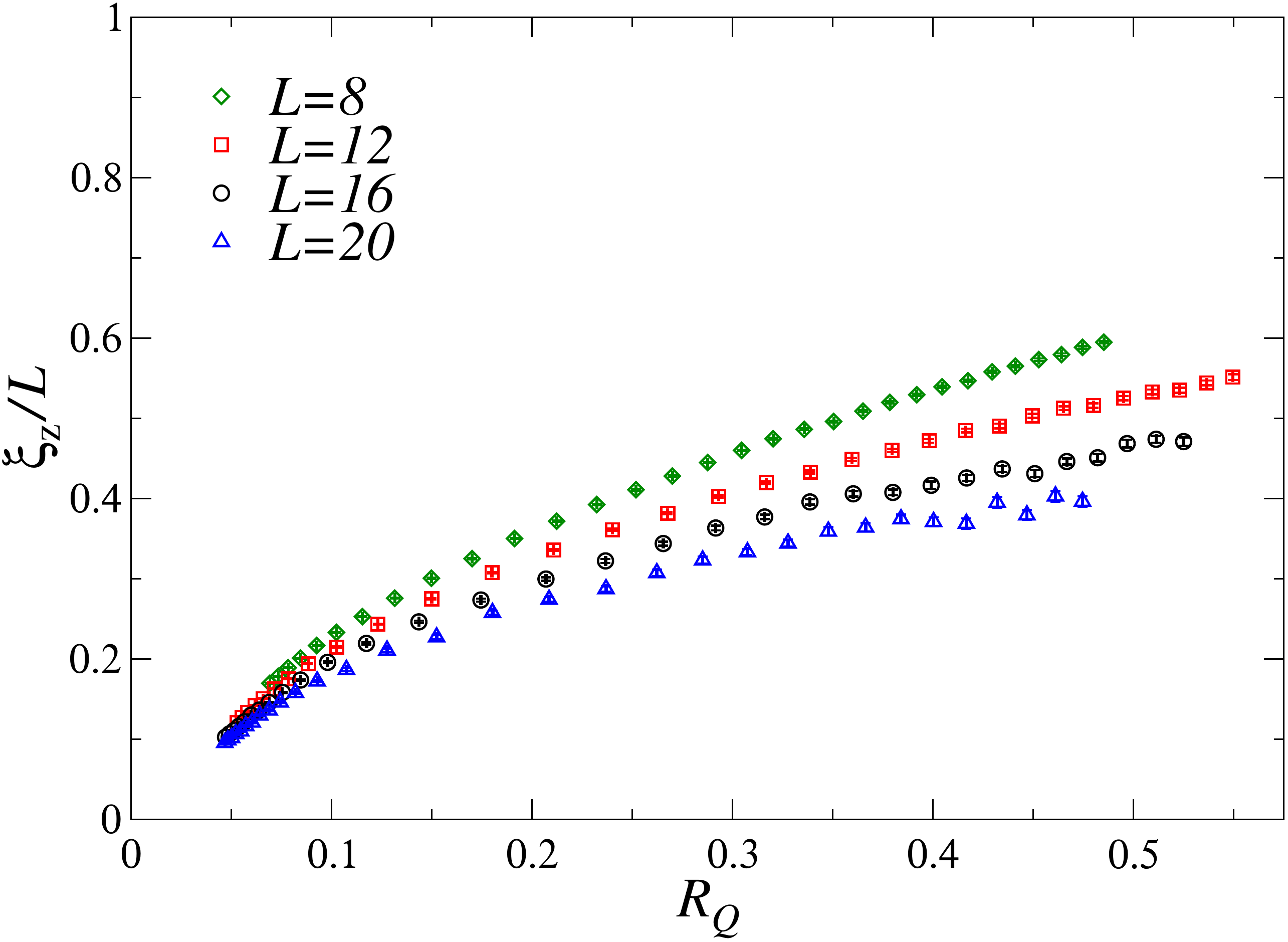}
  \caption{Ratio $\xi_z/L$ for the soft Lorenz gauge fixing with
    $\zeta=5$. Data do not scale, at variance with what happens
    for the hard Lorenz gauge fixing, see
    Fig.~\ref{critzLorenz}. Actually, on the basis of the exact
    results reported in the Appendix, we expect $\xi_z/L \to 0$ for
    $L\to \infty$.}
\label{softdata}
\end{figure}

It is important to remark that scalar fields do not develop any
critical behavior if a soft Lorenz gauge is considered. Indeed, in
this case it is possibe to prove rigorously, see
Refs.~\cite{KK-85,KK-86,BN-87}, App.~C of \cite{Herbut-book} and the
Appendix of this paper, that the scalar-field correlation function
decays exponentially in all phases, with a length scale $\xi_\zeta
\simeq 1/\zeta$, where $\zeta$ is the parameter of the gauge-fixing
term defined in Eq.~(\ref{H-soft}). Numerical results, see
Fig.~\ref{softdata}, confirm the exact predictions.

\begin{figure}
  \includegraphics*[width=0.95\columnwidth]{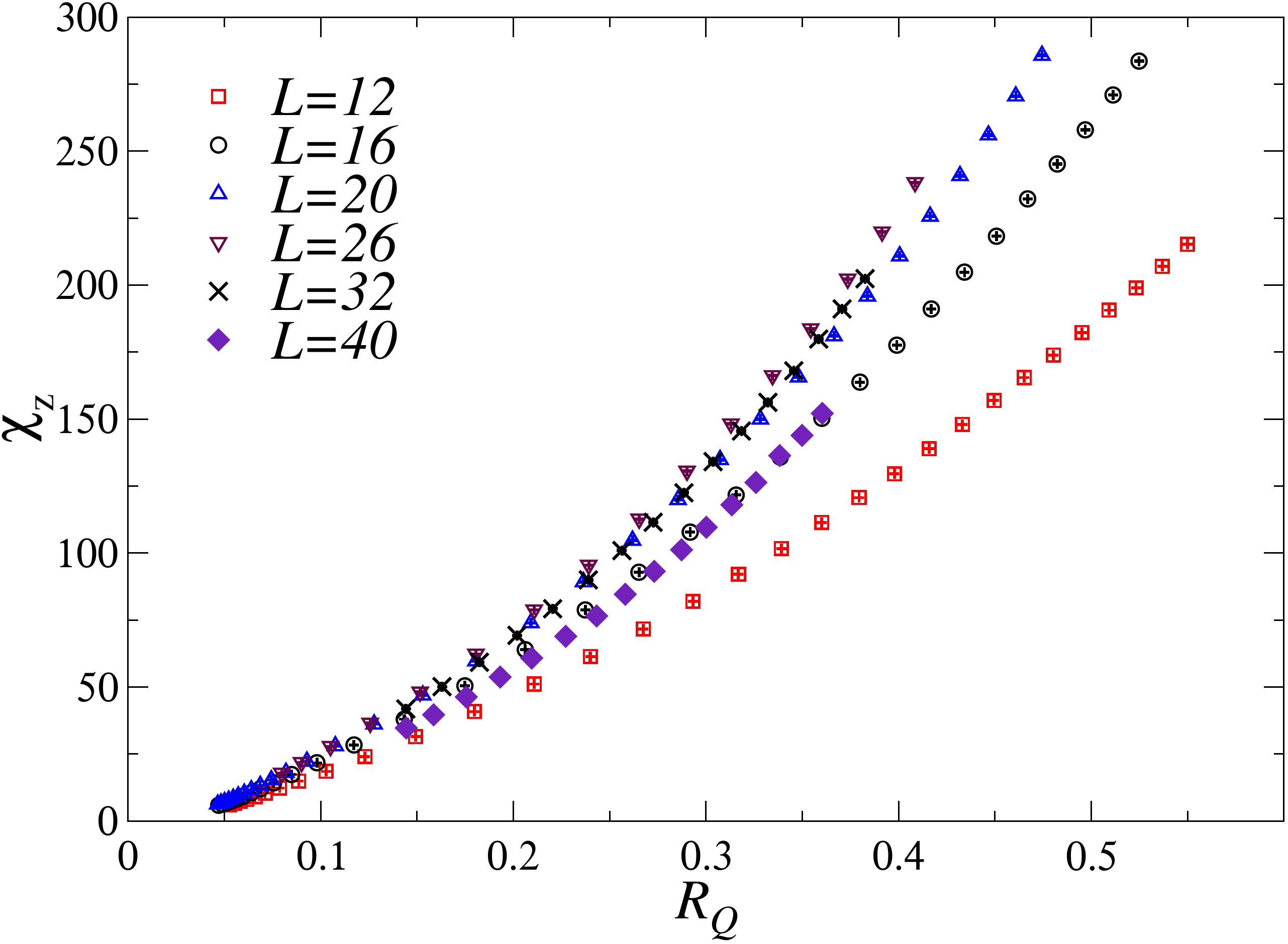}
  \caption{Plot of $\chi_z $ as a function of $R_Q$, for the hard
    axial gauge. Note that the results for $L=40$ are lower than those
    for $L=26,32$ and essentially overlap with those corresponding to
    $L=16$.  }
\label{chiz-axial}
\end{figure}

Finally, let us discuss the hard axial gauge. Numerical results show that 
no long-range order occurs. Indeed, see Fig.~\ref{chiz-axial}, 
$\chi_z$ increases with $L$ for  $L\lesssim 30$, but 
decreases for larger lattices. For instance, the data with $L=40$ 
are significantly smaller than those with $L=32$ and mostly overlap with 
the $L=16$ data. The absence of long-range order is probably due to fact
that, in the infinite-volume limit, the model becomes invariant under a large
class of gauge transformations, namely, those with $\Lambda_{(x,y,z)}$ 
independent of $z$. This large invariance group makes correlations 
between points ${\bm x}_1 = (x_1,y_1,z_1)$ and ${\bm x}_2 = (x_2,y_2,z_2)$
vanish, 
i.e., $\langle \bar{\bm z}_{{\bm x}_1} \cdot {\bm z}_{{\bm x}_2} \rangle = 0$,
unless $x_1 = x_2$ and $y_1 = y_2$. In the Appendix, we prove 
this property in the soft axial gauge, but we expect it to be true 
also in the hard case, being related to a general feature of the axial gauge.

\subsection{Hard-soft crossover in the Lorenz gauge}
\label{scacosoft}

\begin{figure}
  \includegraphics*[width=0.95\columnwidth]{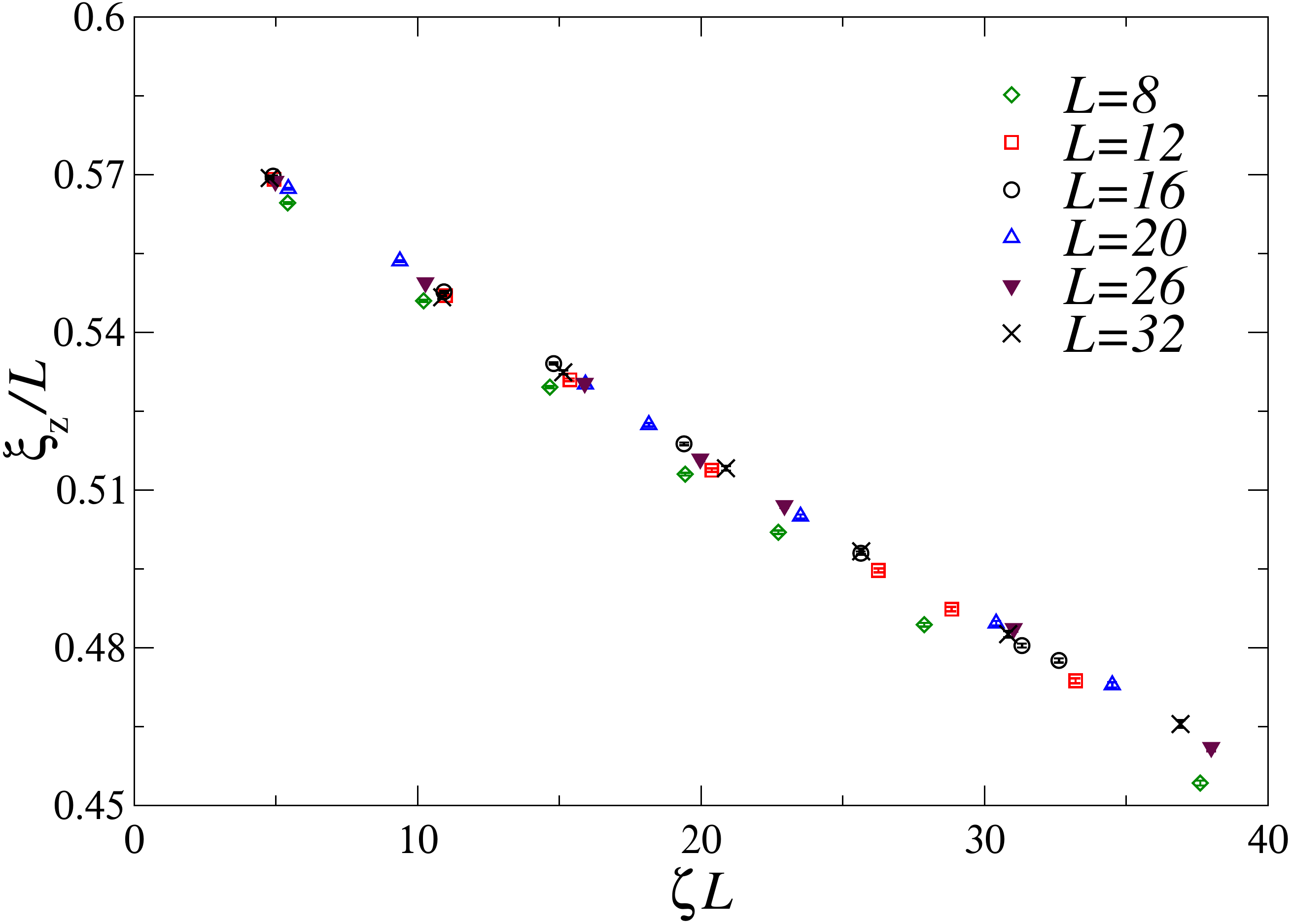}
  \includegraphics*[width=0.95\columnwidth]{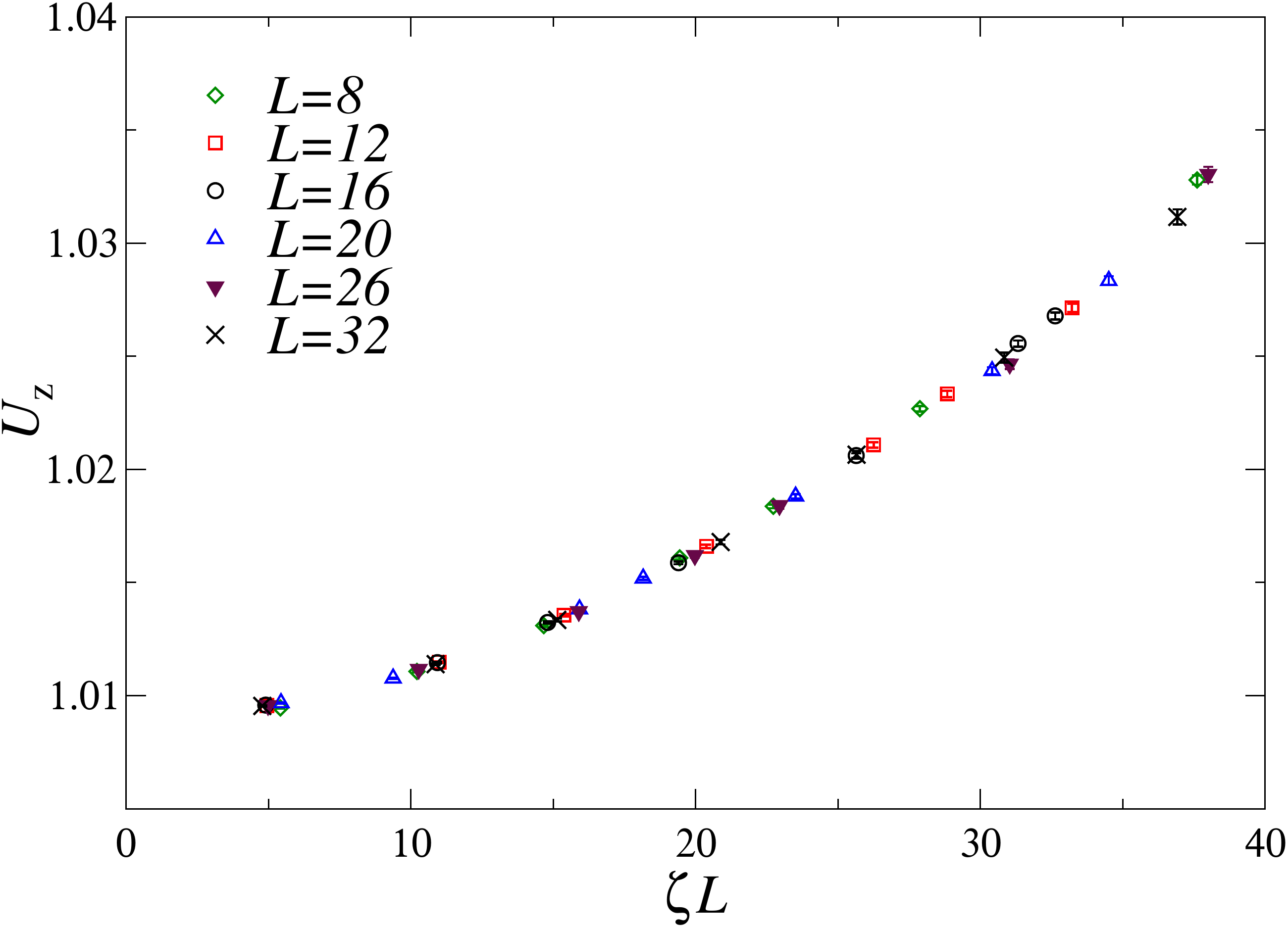}
  \caption{Ratio $\xi_z/L$ (top) and 
    Binder parameter $U_z$ (bottom) at the critical point $J_c$ 
    versus $Z = \zeta L$, in the presence of a soft Lorenz gauge fixing.
    }
\label{crosssoftz}
\end{figure}

As already discussed in Ref.~\cite{BPV-23} and in
Sec.~\ref{scalarcrit}, the behavior of gauge and scalar correlations
differs in the hard and soft implementation of the Lorenz gauge
fixing. If the soft-gauge parameter $\zeta$ is positive, irrespective
of the phase one is considering, scalar-field correlations are
short-ranged while gauge-field correlations show the typical Coulomb
behavior.  As noted in Ref.~\cite{BPV-23}, the value $\zeta = 0$ is
also singled out by the AH field theory. Indeed, as a consequence of
the Ward identities that follow from gauge invariance, the $\beta$
function of the parameter $\zeta$ takes the simple form
\begin{equation}
  \beta_\zeta = - \zeta \eta_A.
  \label{betazeta}
\end{equation}
The value $\zeta = 0$ is therefore a fixed point of the RG
transformations for the theory with a Lorenz gauge fixing. Moreover,
since $\eta_A > 0$, this fixed point is unstable. If we start the RG
flow from $\zeta\not=0$, then $\zeta$ increases towards infinity, so
that the nongauge-invariant modes become unbounded. Close to $\zeta =
0$, field theory predicts a nontrivial crossover behavior in terms of
the variable
\begin{equation}
  Z\equiv\zeta L^{\eta_A} = \zeta L.
  \label{Zdef}
  \end{equation}
Therefore, close to the critical point $J_c$ (we assume that $\kappa$
is fixed), any RG invariant quantity $R$ that is not gauge invariant,
such as $\xi_A/L$, $U_A$, $\xi_z/L$ and $U_z$, is expected to scale as
\begin{eqnarray}
 R(J,\zeta,L) \approx {\cal R}(X,Z),
 \label{rzscazeta}
\end{eqnarray}
where $X=(J-J_c)L^{1/\nu}$. The behavior observed in the hard Lorenz
gauge fixing is recovered in the limit $Z\to 0$. Analogously,
susceptibilities of local nongauge-invariant observables $O_{\bm x}$
would behave as
\begin{eqnarray}
 \chi_O(J,\zeta,L) \approx L^{3-2y_o} {\cal F}(X,Z).
 \label{chiscazeta}
\end{eqnarray}
Note that the scaling variable $Z=\zeta L$ emerges naturally also from
the analysis of the behavior of scalar-field correlations (see
Ref.~\cite{BN-87} and the appendix), since, in the presence of a soft
Lorenz gauge fixing, correlations at the critical point and in the
Higgs phase decay with length scale $\xi_z \simeq 1/\zeta$, so that $Z
= L/\xi_z$ is the natural RG invariant length ratio.

\begin{figure}
  \includegraphics*[width=0.95\columnwidth]{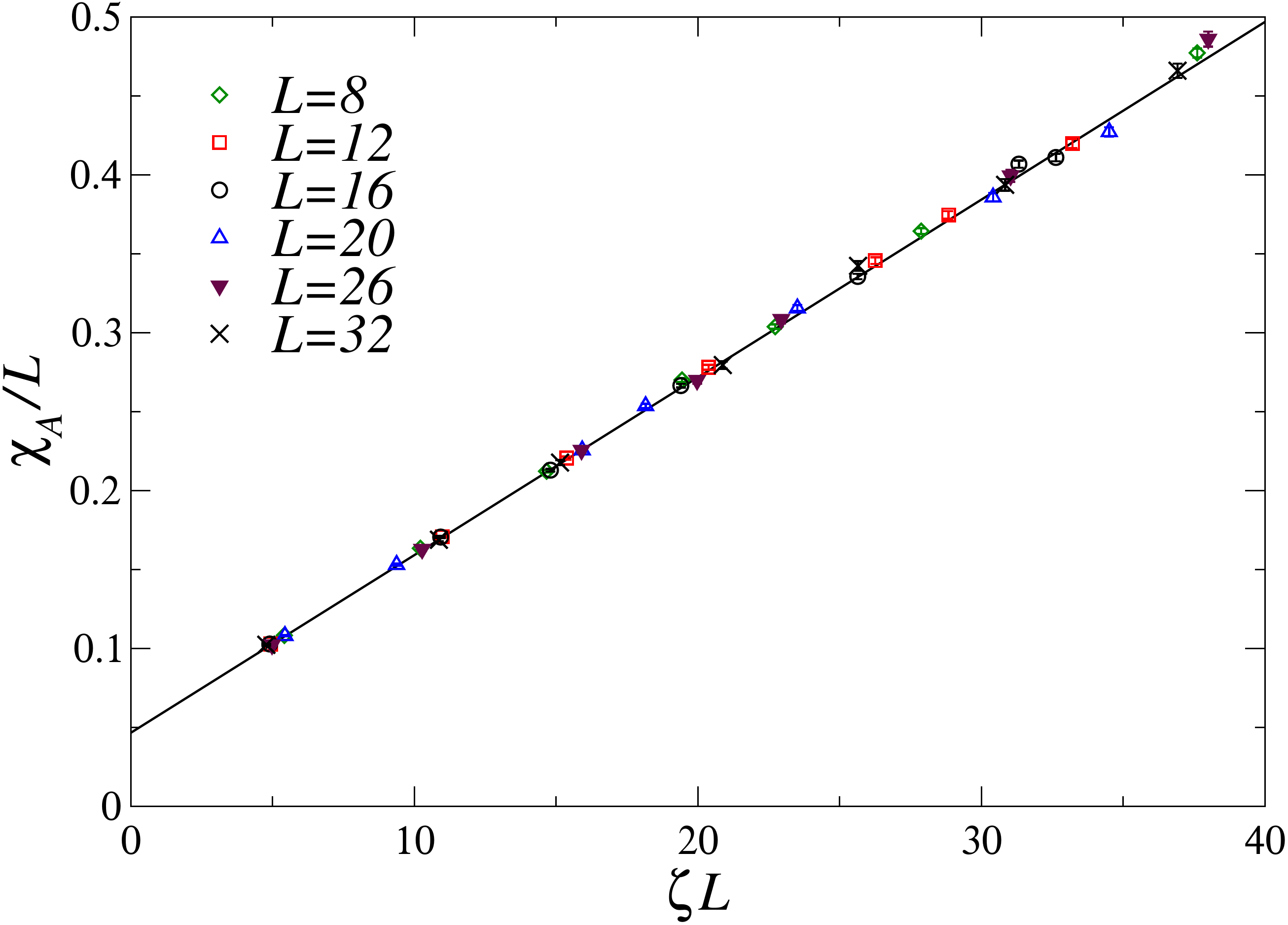}
  \caption{Ratio $\chi_A/L$ at the critical point $J_c$ versus $Z =
    \zeta L$, in the presence of a soft Lorenz gauge fixing.  The line
    is the result of a fit to Eq.~\eqref{crossAsf}, which results in
    $a_0(X=0)\simeq 0.0465$. }
\label{crosssoftA}
\end{figure}  

To verify the critical crossover between the soft and the hard Lorenz
gauge behavior, in Fig.~\ref{crosssoftz} we report $\xi_z/L$ and $U_z$
for several values of $\zeta$ and $L$ at the critical point, i.e., for
$X=0$. Data nicely support the crossover relation~(\ref{rzscazeta}).
The same analysis can be performed for the gauge correlation functions.
We consider the 
susceptibility $\chi_A$, which should scale as in Eq.~~(\ref{chiscazeta}),
with $3-2 y_A = 2 - \eta_A = 1$ and~\cite{BPV-23}
\begin{equation}\label{crossAsf}
\mathcal{F}(X,Z)=a_0(X) + \frac{1}{9\pi^2}Z\ .
\end{equation}
As shown in Fig.~\ref{crosssoftA}, this prediction is fully confirmed by
numerical data.

\section{Conclusions}
\label{conclu}

We have reported a numerical study of 
the 3D lattice AH gauge model with noncompact gauge variables and
multicomponent scalar fields, with Hamiltonian (\ref{AHH}),
along the transition line that separates the Coulomb and Higgs phases.
Previous numerical studies of the
correlation functions of gauge-invariant bilinear scalar
operators~\cite{BPV-21-ncAH,BPV-22}, showed that, for 
$N>N^{\star}$ with $N^{\star}=7(2)$, the CH transitions are continuous and
associated with the stable charged fixed point of the
RG flow of the continuum AH field theory (\ref{AHFT}).

In this paper we extend the previous numerical analyses to the
gauge-dependent correlation functions of the gauge and scalar fields,
to deepen our understanding of the relation between lattice and
field-theoretical gauge-dependent correlation functions, and to
strengthen the identitification of the CH transitions as charged
transitions with an effective AH field-theory description.  For this
purpose we supplement the lattice AH model with a gauge fixing,
considering, in particular, the Lorenz and the axial gauge. They are
implemented as hard constraints on the allowed ensemble
configurations, see Eq.~(\ref{Zhard}).

We study the model with $N=25$ scalar components, for which the
transitions along the CH line are continuous \cite{BPV-21-ncAH}.  The
results in the nonperturbative critical regime show that the
gauge-field correlations have a critical behavior that agrees with
what is predicted by field theory, independently of the gauge fixing
used (this is consistent with the exact results of
Ref.~\cite{BPV-23}).  In particular, the numerical estimate of the
critical exponent $\eta_A$, which is directly related to the RG
dimension of the gauge field, is fully consistent with the exact value
$\eta_A=1$ obtained in the AH field theory with the Lorenz gauge
fixing. More generally, for $d$-dimensional systems the value
$\eta_A=4-d$ can be proved to all orders of perturbation theory using
the Ward-Takahashi identities and assuming the existence of a charged
fixed point~\cite{HT-96}.  The results for the scalar correlations
depend on the gauge fixing. In the hard Lorenz gauge scalar
correlations become critical and we can evaluate the anomalous
dimension of the scalar field, obtaining $\eta_z=-0.069(1)$, which is
not far from the large-$N$ field-theory estimate \cite{HLM-74} $\eta_z
\approx -0.081$. This result provides further evidence of the charged
nature of the fixed point controlling the critical behavior along the
CH transition line for sufficiently large $N$. On the other hand, in
the hard axial gauge scalar correlations are not critical.

We remark that the nontrivial FSS behavior of the scalar-field
correlation function $\langle \bar{\bm z}_{\bm x} \cdot {\bm z}_{\bm
  y} \rangle$ reflects the fact that, in the hard Lorenz gauge, the
scalar field is long-ranged in the Higgs phase, as in the
semiclassical picture usually associated with the Higgs mechanism.  As
proven in Refs.~\cite{KK-85,KK-86,BN-87}, the scalar field does not
order in the Higgs phase in the presence of a soft Lorenz gauge fixing
for any $\zeta>0$, cf. Eq.~(\ref{H-soft}). Therefore, one does not
expect the scalar field to be long-ranged at the critical point for
any $\zeta > 0$.  Our numerical results confirm this prediction. In
the RG language, the hard case $\zeta=0$ can be interpreted as a
multicritical point associated with an unstable RG fixed point. Close
to $\zeta=0$, one therefore expects a universal critical crossover in
terms of the gauge parameter. Numerical data are in full agreement.

We finally mention that a substantially different behavior is expected
along the CM transition line in Fig.~\ref{phdiasketchncLAH}.  Indeed,
the transitions between the Coulomb and the molecular phase are
conjectured to be associated with a Landau-Ginzburg-Wilson $\Phi^4$
theory without gauge
fields~\cite{BPV-21-ncAH,PV-19-CP,PV-19-AH3d,PV-20-largeNCP}, defined
in terms of a tensor order-parameter field $\Phi^{ab} \sim Q_{{\bm
    x}}^{ab}\equiv \bar{z}_{\bm x}^a z_{\bm x}^b -
\delta^{ab}/N$. This allows one to predict first-order transitions for
any $N\ge 3$, and continuous transitions belonging to the 3D O(3)
vector, or CP$^1$, universality class for $N=2$.  Therefore, the
stable fixed point for $N=2$ is {\em uncharged}, i.e., the gauge
fields do not have any relevant role along the critical line, apart
from restricting the critical modes to those that are gauge invariant.
Therefore, even in the presence of a hard Lorenz gauge fixing, we do
not expect gauge-dependent correlations to show any particular scaling
behavior for $N=2$ along the CM line (scalar correlations should be
short-ranged, while gauge correlations should show Coulomb behavior as
for $J=0$), Numerical analyses for $N=2$ along the CM line, that we do
not report, definitely confirm this scenario.

\appendix

\section{Some results on soft gauges} 

In this appendix we wish to relate the expressions of 
the correlation functions in soft and hard gauges, generalizing the 
results of Refs.~\cite{KK-85,KK-86,BN-87,Herbut-book}. We will consider 
the model in $d$ dimensions, as many of the results are not specific 
of the three-dimensional model.
We consider
a generic translation-invariant gauge fixing function that is linear in 
the gauge fields:
\begin{equation}
  F_{\bm x} (A) = \sum_{{\bm y}\mu} M_{{\bm x}-{\bm y},\mu} 
     A_{{\bm y},\mu}.
\end{equation}
Under a gauge transformation (\ref{gautra}) we have
\begin{equation}
   F_{\bm x} (A') = F_{\bm x} (A) + f_{\bm x} (\Lambda),
\end{equation}
with 
\begin{equation}
   f_{\bm x} (\Lambda) = -\sum_{{\bm y}\mu} M_{{\bm x}-{\bm y},\mu}
      \Delta_\mu \Lambda_{\bm x},
\end{equation}
where $\Delta_\mu f_{\bm x} = f_{{\bm x}+\hat\mu} - f_{\bm x}$ is a
lattice derivative.  We assume the gauge fixing to be complete so that
the integral $I$ defined by
\begin{equation}
   I = \int [d\Lambda] \prod_{{\bm x}} 
        \delta(F_{\bm x} (A) + f_{\bm x} (\Lambda)) 
\end{equation}
is strictly positive.  We now consider a correlation function of the
scalar field and proceed as outlined in Ref.~\cite{BN-87}. Focusing
for simplicity on the two-point correlation function $\langle
{\bar{z}}_{\bm x} z_{\bm y} \rangle_\zeta$ in a soft gauge with
parameter $\zeta$, we write
\begin{eqnarray}
&& Z_{\zeta} \langle {\bar{z}}_{\bm x} z_{\bm y} \rangle_\zeta =  \int
[dA] [dz d\bar{z}] e^{-H - H_{\rm gf}} {\bar{z}}_{\bm x} z_{\bm y}
\times  \nonumber \\
&&\qquad \times {1\over I} \int [d\Lambda] \prod_{{\bm
    w}} \delta(F_{\bm w} (A) + f_{\bm w} (\Lambda)),
\end{eqnarray}
where $Z_{\zeta}$ is the normalizing partition function. Then, 
we perform the change of 
variables $A_{{\bm x},\mu}' = A_{{\bm x},\mu} - \Delta_\mu \Lambda_{\bm x}$
 and $z_{\bm x}' = \exp(i \Lambda_{\bm x}) z_{\bm x} $ obtaining
\begin{eqnarray}
&& Z_{\zeta} \langle {\bar{z}}_{\bm x} z_{\bm y} \rangle = 
    {1 \over I} \int [dA'] [dz' d\bar{z}'] [d\Lambda] 
       e^{-H - {1\over 2\zeta} \sum_{\bm w} f_w(\Lambda)^2} 
\nonumber \\ && \qquad \times \;
     {\bar{z}}_{\bm x}' z_{\bm y}' \prod_{\bm w} \delta(F_{\bm w} (A')) 
        e^{i\Lambda_{\bm x} - i\Lambda_{\bm y}}
 \\
   && \quad = {1 \over I} Z_H \langle {\bar{z}}_{\bm x} z_{\bm y} \rangle_H 
    \int [d\Lambda] e^{- {1\over 2\zeta} \sum_{\bm w} f_w(\Lambda)^2}
     e^{i\Lambda_{\bm x} - i\Lambda_{\bm y}},
\nonumber 
\end{eqnarray}
where $Z_H$ and $\langle \cdot \rangle_H$ are the partition function
and the average for the model with the hard gauge fixing, respectively.
By repeating the same steps for the partition function we obtain 
a similar relation betweem $Z_{\zeta}$ and $Z_H$. We end up with 
\begin{equation}
     \langle {\bar{z}}_{\bm x} z_{\bm y} \rangle_\zeta = 
     R_{{\bm x},{\bm y}} \langle {\bar{z}}_{\bm x} z_{\bm y} \rangle_H 
\end{equation}
with 
\begin{equation}
R_{{\bm x},{\bm y}} = 
     {\int [d\Lambda] e^{- {1\over 2\zeta} \sum_{\bm w} f_w(\Lambda)^2}
     e^{i\Lambda_{\bm x} - i\Lambda_{\bm y}} \over 
    \int [d\Lambda] e^{- {1\over 2\zeta} \sum_{\bm w} f_w(\Lambda)^2} }
\end{equation}
This quantity is a Gaussian integral which gives 
\begin{equation}
R_{{\bm x},{\bm y}} = \exp\Bigl(
  -{{\zeta}  I_{{\bm x},{\bm y}}} \Bigr)
\end{equation}
where, for translation-invariant boundary conditions
\begin{equation}
I_{{\bm x},{\bm y}} = {1\over L^d} 
   \sum_{\bm p} {1 - e^{-i {\bm p} \cdot ({\bm x} - {\bm y})}  \over 
          D(p)} ,
\end{equation}
with 
\begin{equation}
    D(p) = \left| \sum_\mu M_\mu({\bm p}) \hat{p}_\mu e^{-ip_\mu/2} 
   \right|^2 
\end{equation}
and $\hat{p}_\mu = 2 \sin (p_\mu/2)$.

\subsection{Lorenz gauge}

In the Lorenz gauge we have 
\begin{equation}
I_{{\bm x},{\bm y}} = I({\bm x} - \bm{y}) = {1\over L^d}
   \sum_{\bm p} {1 - e^{-i {\bm p} \cdot ({\bm x} - {\bm y})}  \over
     ({\hat p}^2)^2}
\end{equation}
where $\hat{p}^2 = \sum_\mu \hat{p}_\mu^2$.  In the infinite-volume
limit, for $2 < d < 4$, the asymptotic large-distance behavior can be
easily computed using the representation (we assume that $x_\mu \ge 0$
for any $\mu$):
\begin{equation}
I({\bm x}) = \int_0^\infty \alpha d\alpha\, e^{-2 d\alpha} 
    \left[I_0(2 \alpha)^d - \prod_{\mu} I_{x_\mu}(2\alpha)\right].
\end{equation}
For $x_\mu \to \infty$ for all values of $\mu$ we can 
approximate the integral with  ($|{\bm x}|^2 = \sum_\mu x_\mu^2$)
\begin{equation}
I({\bm x}) \approx \int_0^\infty \alpha d\alpha\, (4 \pi \alpha)^{-d/2} 
    (1 - e^{-|{\bm x}|^2/(4 \alpha)} ), 
\end{equation}
which implies 
\begin{equation}
I({\bm x}) \approx - (4 \pi)^{-d/2} \Gamma(d/2-2) (|{\bm x}|/2)^{4-d}.
\end{equation}
In three dimensions $I({\bm x}) = |{\bm x}|/(8 \pi)$, so that 
\begin{equation}
R_{{\bm x},{\bm y}} = \exp\Bigl(-{\zeta\over 8\pi} 
     |{\bm x}-{\bm y}|\Bigr).
\end{equation}
Therefore, in the soft Lorenz gauge, the correlation function is
short-ranged in all phases, as already noted in
Ref.~\cite{Herbut-book}.  At the critical point the scalar correlation
function with hard Lorenz gauge fixing decreases as a power of the
distance (more precisely, as $1/|{\bm x}|^{1+\eta_z}$), while in the
Higgs phase it converges to a constant for large distances. Thus, in
these two cases the soft-gauge correlation function decays
exponentially with correlation length $\xi_\zeta = 8 \pi/\zeta$.

In four dimensions, the large-$|x|$ expansion of $I({\bm x})$
has been computed in Ref.~\cite{LW-95}, obtaining
\begin{equation}
I({\bm x}) \approx {1\over 8 \pi^2} \log|{\bm x}| + c + O(|{\bm x}|^{-2}), 
\end{equation}
with $c\approx 0.0225449$.
It follows that, at large distances, the soft and hard correlation
function are related by
\begin{equation}
\langle {\bar{z}}_{\bm x} z_{\bm y} \rangle_\zeta =
     e^{- c\,\zeta/2} \langle {\bar{z}}_{\bm x} z_{\bm y} \rangle_H 
      |x|^{-\zeta/(8\pi^2)} .
\end{equation}
This relation gives the complete $\zeta$-dependence of 
the large-distance behavior of 
the correlation function and therefore also of the anomalous dimension 
$\eta_z$ of the field. It implies 
\begin{equation}
   \eta_z(\zeta) = \eta_{zH} + {\zeta \over 8\pi^2},
\end{equation}
where $\eta_{zH}$ is the value of the exponent in the hard gauge.
This result has already been derived in perturbation theory, see, e.g., 
Ref.~\cite{ZJ-book} (we should note that $\zeta$ as defined here differs
by a factor $e^2$, $e$ being the electric charge, from the quantity
used in perturbative field theory). 
The method used here provides an alternative
nonperturbative derivation.

\subsection{Axial gauge} 
     
In the axial gauge we have 
\begin{equation}
I_{{\bm x},{\bm y}} = I({\bm x} - \bm{y}) = {1\over L^d}
   \sum_{\bm p} {1 - e^{-i {\bm p} \cdot ({\bm x} - {\bm y})}  \over
     {\hat p}^2_d} .
\end{equation}
To compute $I({\bm x})$, we write ${\bm x} = ({\bm x}_T,x_d)$. 
If ${\bm x}_T\not = 0$, we have simply
\begin{equation}
    I({\bm x}) = {1\over L^d}  \sum_{\bm p} {1\over \hat{p}_d^2}  = 
        {1\over L} \sum_p {1\over \hat{p}^2} = {L\over 4}.
\end{equation}
On the other hand, for ${\bm x}_T = 0$, we have 
\begin{equation}
    I({\bm x}) = D_1(x_d) = {1\over L} \sum_p {1 - e^{ip x_d} \over \hat{p}^2}.
\end{equation}
This one-dimensional sum can be easily computed. It is enough to note that 
\begin{eqnarray}
D_1(x+1) - 2 D_1(x) + D_1(x-1) &=& {1\over L} \sum_p e^{ipx} 
\\ &= &
    \sum_k (-1)^k \delta_{x,kL}, \nonumber 
\end{eqnarray}
where in the last step we have assumed antiperiodic boundary conditions. 
This recursion relation can be easily solved obtaining 
\begin{equation}
   D_1(x) = {|x|\over2} 
\end{equation}
in $x\in [-L,L]$. 
Therefore, we have 
\begin{eqnarray}
R_{{\bm x},{\bm y}}=e^{-\zeta L/4} \hphantom{????!}
    &&   \hbox{if ${\bm x}_T\not= {\bm y}_T$} 
\\
R_{{\bm x},{\bm y}} = e^{-\zeta|x_d-y_d|/2}    &&   \hbox{if ${\bm x}_T={\bm
y}_T$.}
\nonumber 
\end{eqnarray}
Thus, in a soft gauge, the infinite-volume correlation function 
$\langle \bar{z}_{\bm x} z_{\bm y} \rangle_\zeta$ vanishes if 
${\bm x}_T\not= {\bm y}_T$ for all values of $\zeta$, while it decays
exponentially for ${\bm x}_T = {\bm y}_T$. The corresponding scalar
suceptibility $\chi_z$ is therefore finite for any $\zeta$. 
Note also that $R_{{\bm x},{\bm y}}$ decreases with $L$, and therefore
we expect that $\chi_z$ decreases with increasing the volume.
As discussed in Ref.~\cite{BPV-23}, in the axial case the parameter $\zeta$ 
appears to be irrelevant: gauge correlations have the same behavior for any
$\zeta$, including $\zeta = 0$. We thus expect the correlation function
to have the same qualitative behavior also in the hard-axial gauge, i.e. that 
it vanishes for ${\bm x}_T\not= {\bm y}_T$ in the infinite-volume limit.
Therefore, also in the hard axial gauge 
no critical behavior is expected in the scalar sector. The results 
presented in Sec.~\ref{scalarcrit} confirm this prediction.

\end{document}